\documentclass[%
superscriptaddress, %
twocolumn, %
showpacs,
 amsmath,amssymb,
 aps,
prb,
floatfix, longbibliography ]{revtex4-1}
\usepackage{graphicx}
\usepackage{dcolumn}
\usepackage{bm}
\usepackage{soul,color}

\usepackage[colorlinks,linkcolor=blue,anchorcolor=blue,citecolor=blue,urlcolor=blue]{hyperref}
\usepackage{tabularx}
\usepackage{mathtools}
\usepackage{amsmath}

\newcommand{\mb}[1]{\mathbf{#1}}
\newcommand{\mr}[1]{\mathrm{#1}}

\begin{document}

\title{Collective excitations of the Chern-insulator states in commensurate double moir\'{e}
superlattices of twisted bilayer graphene on hexagonal boron nitride}

\author{Xianqing Lin}
\email[E-mail: ]{xqlin@zjut.edu.cn}
\affiliation{College of Science,
             Zhejiang University of Technology,
             Hangzhou 310023, China}

\author{Quan Zhou}

\affiliation{College of Science,
             Zhejiang University of Technology,
             Hangzhou 310023, China}

\author{Cheng Li}

\affiliation{College of Science,
             Zhejiang University of Technology,
             Hangzhou 310023, China}

\author{Jun Ni}
\affiliation{State Key Laboratory of Low-Dimensional Quantum Physics and Frontier Science Center for Quantum Information,
             Department of Physics, Tsinghua University, Beijing 100084,
             China}

\date{\today}

\begin{abstract}
We study the collective excitation modes of the Chern insulator states
in magic-angle twisted bilayer graphene aligned with hexagonal boron nitride (TBG/BN) at odd
integer fillings ($\nu$) of the flat bands.
For the $1 \times 1$ commensurate double moir\'{e} superlattices in TBG/BN at three twist
angles ($\theta'$) between BN and graphene, self-consistent Hartree-Fock calculations
show that the electron-electron interaction and the broken $C_{2z}$ symmetry lead to the
Chern-insulator ground states with valley-spin flavor polarized HF bands at odd $\nu$.
In the active-band approximation, the HF bands in the same flavor of TBG/BN are much more
separated than those of the pristine TBG with TBG/BN having a larger intra-flavor band gap so that
the energies of the lowest intra-flavor exciton modes of TBG/BN computed within the
time-dependent HF method are much higher than those of TBG and reach about 20 meV,
and the exciton wavefunctions of TBG/BN become less localized than those of TBG.
The inter-flavor valley-wave modes in TBG/BN have excitation energies higher
than 2.5 meV which is also much larger than that of TBG,
while the spin-wave modes all have zero excitation gap.
In contrast to TBG with particle-hole symmetric excitation modes for positive and
negative $\nu$, the excitation spectrums and gaps of TBG/BN at positive $\nu$ are
rather different from those at negative $\nu$. The quantitative behavior of the excitation
spectrum of TBG/BN also varies with $\theta'$.
Full HF calculations demonstrate that more HF bands besides the two central bands can
have rather large contributions from the single-particle flat-band states,
then the lowest exciton modes that determine the optical properties of the Chern
insulator states in TBG/BN are generally the ones between the remote and flat-like bands,
while the valley-wave modes have similar energies as those in the active-band approximation.
\end{abstract}

\pacs{%
}



\maketitle


\section{Introduction}

\begin{figure*}[t]
\begin{center}
\includegraphics[width=2.0\columnwidth]{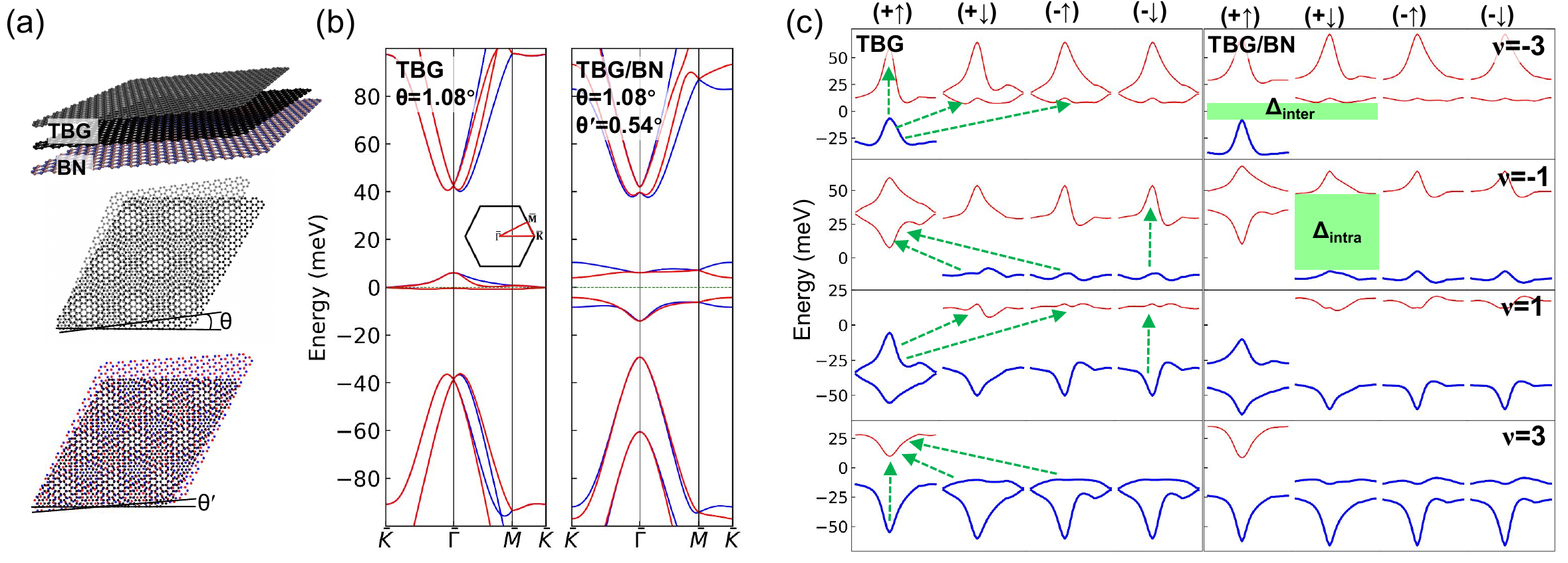}
\end{center}
\caption{(Color online)
The single-particle and HF band structures of TBG and TBG/BN.
(a) The schematic view of the commensurate double moir\'{e} superlattices in TBG/BN.
The twist angle ($\theta$) between the graphene layers and that ($\theta'$) between
graphene and BN are labeled.
(b) The single-particle band structures of TBG at the magic $\theta = 1.08^{\circ}$
and a commensurate supercell of TBG/BN at $\theta' = 0.54^{\circ}$.
The red and blue lines represent bands in the $\xi = +$ valley and
the $\xi = -$ valley, respectively.
(c) The HF band structures of the Chern-insulator ground states at odd $\nu$ in
the active-band approximation for the TBG and TBG/BN systems in (b).
The HF bands of each flavor are plotted separately along the k-point path same as that in
(b), and the flavor is labeled by the valley (+ or -) and spin ($\uparrow$ or $\downarrow$) indices.
The conduction and valence bands are represented by the red and blue lines, respectively.
The intra-flavor energy gap ($\Delta_{intra}$) between the conduction
and valence bands in the same flavor and
the inter-favor energy gap ($\Delta_{inter}$) between the highest valence band
in one flavor and the lowest conduction band in another flavor
are labeled by the green shades. The spin-wave excitation between the valence
and conduction bands in two different flavors with the same
valley index but the opposite spin indices, and the valley-wave excitation
between the bands in two flavors with the same spin index but the opposite valley indices,
and the exciton excitation between bands in the same flavor are
indicated by the dashed green lines.
\label{fig1}}
\end{figure*}

\begin{figure*}[t]
\begin{center}
\includegraphics[width=1.6\columnwidth]{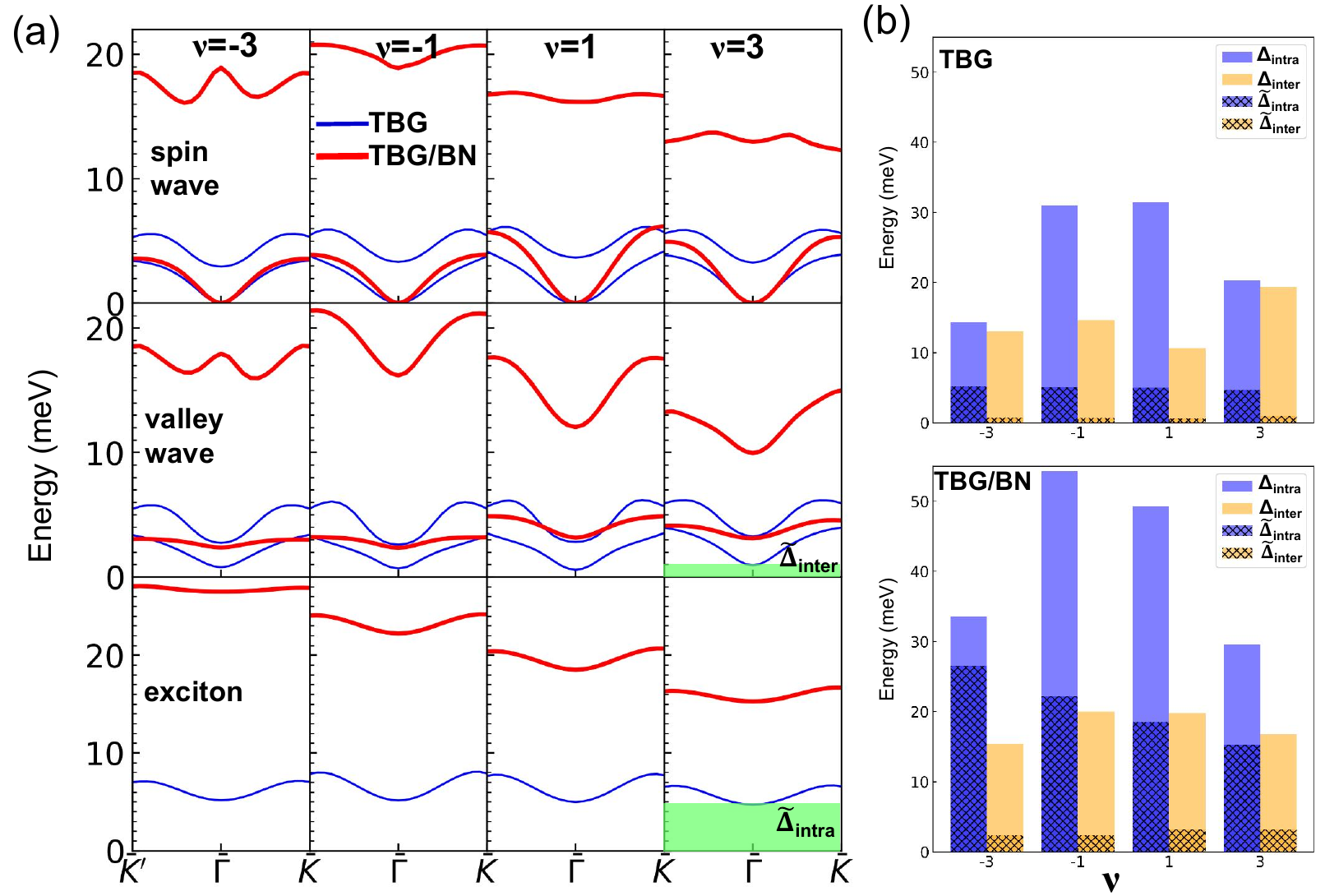}
\end{center}
\caption{(Color online)
The collective excitation modes of TBG/BN and the pristine TBG at odd $\nu$.
(a) The energy spectrums of the two or one lowest excitation modes as a function of
the wave vector $\mb{q}$ for the spin-wave, valley-wave and exciton excitations.
The blue and red lines represent the excitation bands of the pristine TBG and
the TBG/BN at $\theta' = 0.54^{\circ}$, respectively.
$\bar{K}'$ in the k-point path is just the opposite of $\bar{K}$.
The valley-wave excitation gap ($\tilde{\Delta}_{inter}$) and the exciton
excitation gap ($\tilde{\Delta}_{intra}$) at $\mb{q} = \mb{0}$ are labeled by the green shades.
(b) The HF band gaps and the corresponding excitation gaps at different $\nu$ for
TBG/BN and TBG.
\label{fig2}}
\end{figure*}

\begin{figure}[t]
\begin{center}
\includegraphics[width=1.0\columnwidth]{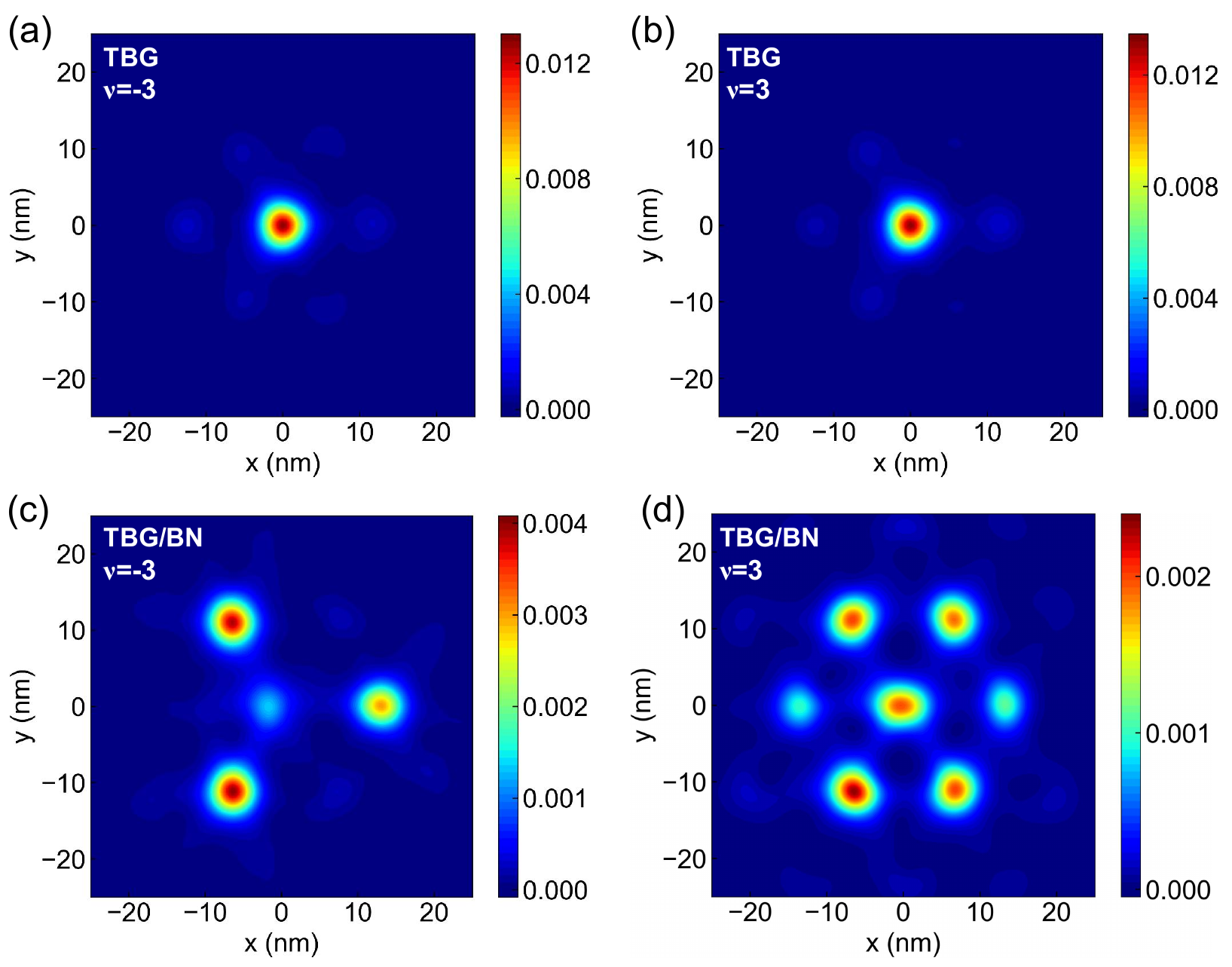}
\end{center}
\caption{(Color online)
The spatial distribution ($|\Psi(\mb{r}_e, \mb{r}_h)|^2$) of the exciton
wavefunction of the lowest mode at $\mb{q} = \mb{0}$ as a function of
the electron position $\mb{r}_e$
with the hole position $\mb{r}_h$ at the origin of a supercell
where the bilayer is locally AA-stacked for TBG at $\nu = -3$ (a), $\nu = 3$ (b), and
TBG/BN with $\theta' = 0.54^{\circ}$ at $\nu = -3$ (c), $\nu = 3$ (d).
\label{fig3}}
\end{figure}

\begin{figure}[t]
\begin{center}
\includegraphics[width=0.9\columnwidth]{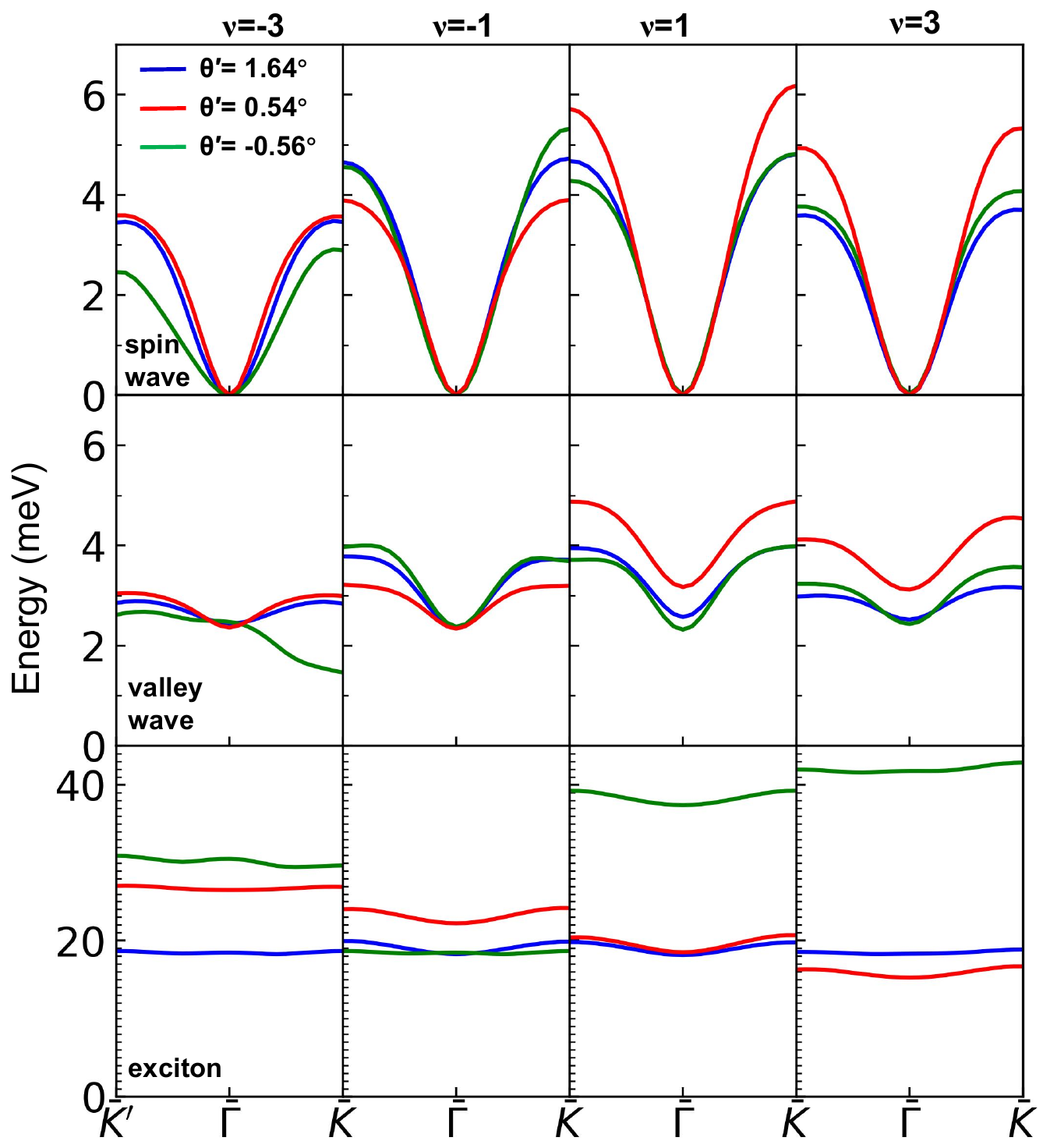}
\end{center}
\caption{(Color online)
The energy spectrums of the lowest spin-wave, valley-wave and exciton
excitation modes at different $\nu$ for
TBG/BN with $\theta' = 1.64^{\circ}$, $0.54^{\circ}$, and $-0.56^{\circ}$.
\label{fig4}}
\end{figure}

\begin{figure}[t]
\begin{center}
\includegraphics[width=0.92\columnwidth]{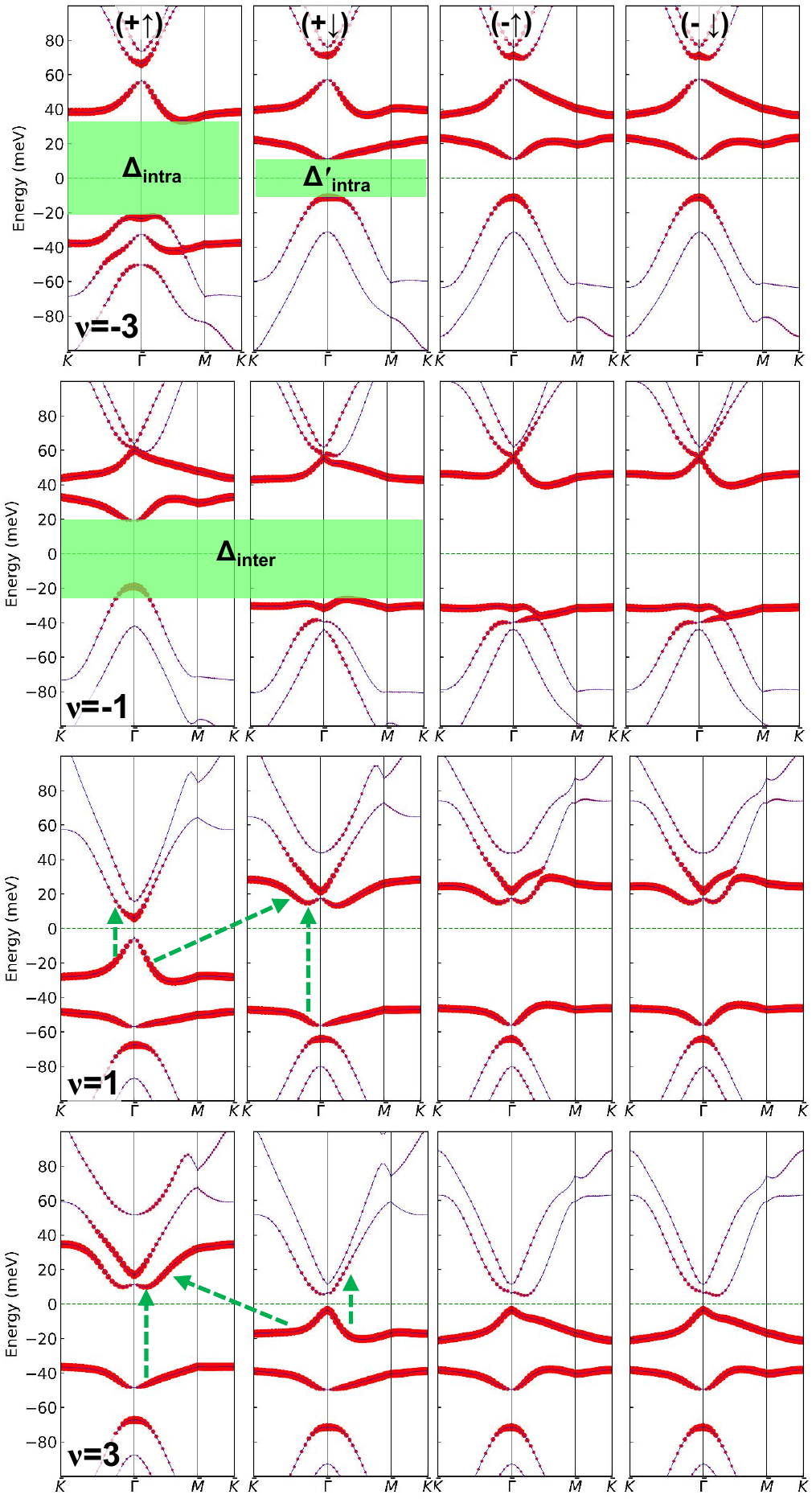}
\end{center}
\caption{(Color online)
The full HF band structures of TBG/BN at $\theta' = 0.54^{\circ}$ for each flavor at different
$\nu$. The magnitude of the projection of each HF band state on the single-particle flat bands
is represented by the size of the red circle. The band gap $\Delta_{intra}$ between
the intra-flavor flat-like bands, the $\Delta_{inter}$ between the inter-flavor flat-like bands,
and the $\Delta'_{intra}$ between the remote and flat-like bands are labeled by the green shades.
The inter-flavor valley-wave excitation mainly between the flat-like bands,
the intra-flavor exciton excitation mainly between the flat-like bands, and the
intra-flavor exciton excitation mainly between the flat-like bands and the remote bands are
indicated by the dashed green lines.
\label{fig5}}
\end{figure}

\begin{figure*}[t]
\begin{center}
\includegraphics[width=1.8\columnwidth]{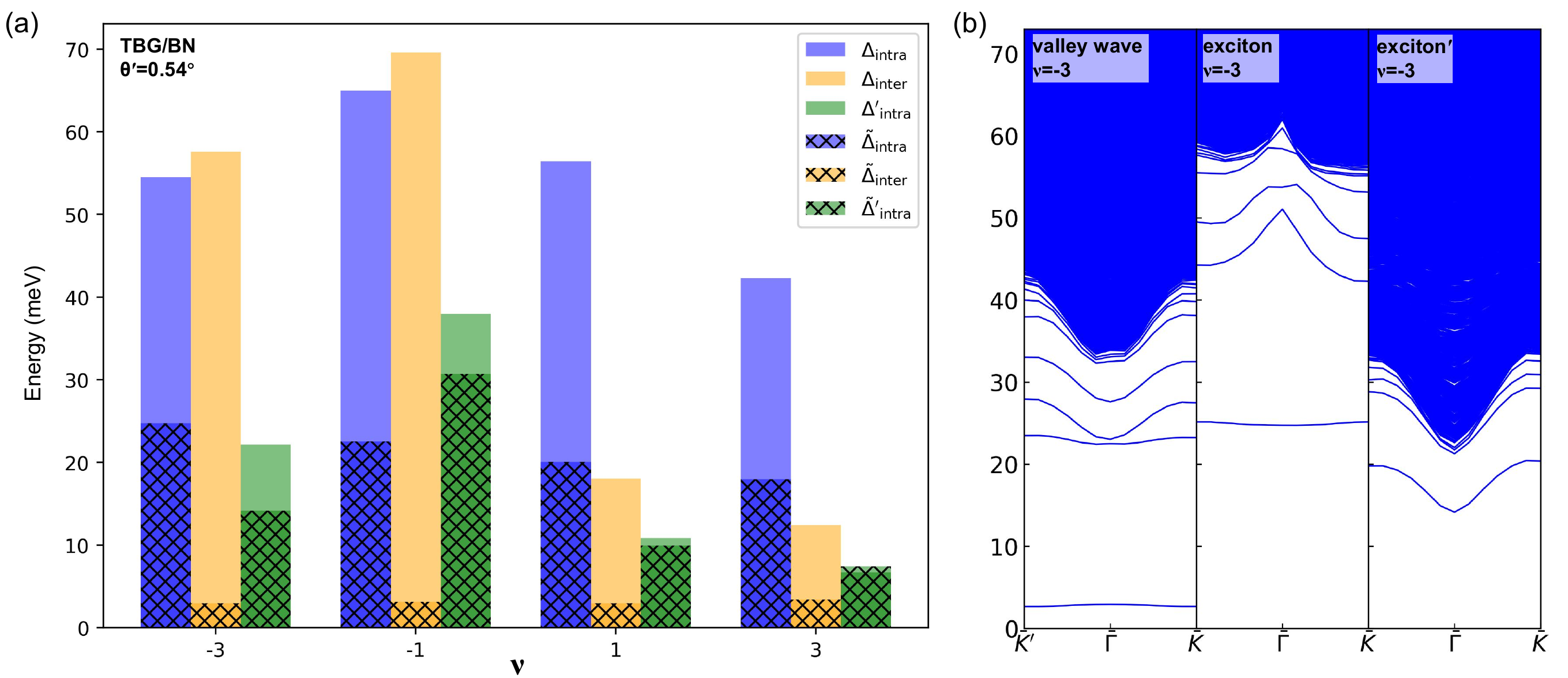}
\end{center}
\caption{(Color online)
(a) The band gaps $\Delta_{intra}$, $\Delta_{inter}$, and $\Delta'_{intra}$ and the corresponding
excitation gaps $\tilde{\Delta}_{intra}$, $\tilde{\Delta}_{inter}$, and $\tilde{\Delta}'_{intra}$
from the full HF calculations at different $\nu$ for TBG/BN with $\theta' = 0.54^{\circ}$.
(b) The energy spectrums for the
valley-wave excitation, the exciton excitation mainly between flat-like bands, and the
exciton excitation (dented by exciton$'$) mainly between empty flat-like bands and the filled
remote bands at $\nu = -3$.
\label{fig6}}
\end{figure*}

Flat bands with vanishing band widths and well separated from other
remote bands occur around the Fermi level in magic-angle twisted bilayer
graphene (TBG)\cite{Bistritzer12233,Morell2010,LopesdosSantos2012,Fang2016,
Carr2017,OriginPhysRevLett.122.106405,WKB-Estimate-2021},
and the experimental realization of such TBG intrigued great
interest in exploring various electronic, transport and optical properties
associated with the flat bands\cite{Cao2018,cao2018unconventional,Kim3364,lu2019superconductors,xie2019spectroscopic,
Maximized-electron-2019,Charge-order-2019,Electronic-correlations-2019,TuningYankowitz1059,
uri2020mapping,Strongly-correlated-2020,Cascade-of-e-2020,Cascade-of-p-2020,
Independent-superconductors-2020,Untying-the-2020,In-situ-2020,Entropic-evidence-2021,
Observation-of-2021,Intelligent-infrared-2022,
Electronic-structure-methods-2020,Graphene-bilayers-2020,The-marvels-2021,Orbital-magnetic-2021}.
The emergence of correlated insulator states at integer filling of the flat bands
in TBG and the superconductivity in the vicinity of these insulating states
have been observed and theoretically
comprehended\cite{Cao2018,cao2018unconventional,Kim3364,lu2019superconductors,xie2019spectroscopic,
Maximized-electron-2019,Charge-order-2019,Electronic-correlations-2019,TuningYankowitz1059,
uri2020mapping,Strongly-correlated-2020,Cascade-of-e-2020,
Cascade-of-p-2020,Independent-superconductors-2020,Untying-the-2020,Entropic-evidence-2021,
Origin-of-mott2018,Maximally-Localized-2018,Symmetry-Maximally-2018,
Unconventional-Supercond-2018,Chiral-Spin-2018,Strong-Coupl-2019,
Bridging-Hubbard-2019,Gate-TunableTopological-2019,Twisted209zhang,Ferromagnetism-in-2020,Nature2020xie,Ground-State-2020,Mechanism2020nick,
Correlated-insulating-2020,Non-AbelianDirac-2020,Collective-Excitations-2020,Theories-for-2021,Hybrid-Wannier-2021,
Nematic-topological-2021,Correlation-InducedInsulating-2021,
Symmetry-breaking-2020lin,Twisted-bilayeriii-2021,Twisted-bilayeriv-2021,Twisted-bilayerv-2021,Kekul-Spiral-2021}.
There are eight single-particle flat bands taking into account the spin and valley
degrees in TBG, then the electron filling of the flat bands
per moir\'{e} supercell relative to charge neutrality point (CNP) is in the range of -4 to 4.
At odd $\nu$, the ground states are Chern insulators with spontaneously broken symmetry
in the valley and spin degrees due to the electron-electron (e-e) interaction
\cite{Mechanism2020nick,Twisted209zhang,Ferromagnetism-in-2020,Collective-Excitations-2020,Nature2020xie,
Theories-for-2021,Hybrid-Wannier-2021,Correlation-InducedInsulating-2021,Symmetry-breaking-2020lin,Kekul-Spiral-2021}.
The alignment of TBG with BN breaks the $C_{2z}$ symmetry in the relaxed atomic
structure and the single-particle Hamiltonian\cite{Symmetry-breaking-2020lin,Band-structure-2020,Misalignment2021Lin,
Moire2021Shi,Quasiperiodicity2021Mao,Electron-hole-asymmetry-2021,An-atomistic-2022}
and thus enhances the energy gaps
of such Chern insulator states\cite{Twisted209zhang,Mechanism2020nick,Collective-Excitations-2020,
Symmetry-breaking-2020lin}.
In particular, the quantum anomalous Hall effect associated with their finite Chern numbers
has been experimentally realized in TBG aligned with BN (TBG/BN)\cite{Intrinsic2020Serlin,Imaging-orbital-2021}.
For such insulating correlated states, low-energy collective excitation states may
appear within the gap due to the Coulomb interaction between the particle and hole states.
In experiments, the observed Pomeranchuk effect from the measured electron
compressibility in TBG at extremely low temperatures implies the presence of the
low-energy collective excitations for the correlated insulator states\cite{Entropic-evidence-2021}.
The optical excitations in the infrared regime have also been observed in twisted
graphene systems around the integer fillings of
the flat bands\cite{Observation-of-2021,Intelligent-infrared-2022,In-situ-2020}.

For the pristine TBG or the TBG with a sublattice potential
difference\cite{Collective-Excitations-2020,Twisted-bilayeriv-2021,Twisted-bilayerv-2021,
Lattice-collective-2021},
theoretical analysis or Hartree-Fock (HF) calculations indeed demonstrated
the occurrence of the low-energy collective
excitation modes of inter-flavor spin wave, valley wave and intra-flavor exciton at odd $\nu$.
The spin-wave excitation states are Goldstone modes with a zero
excitation gap\cite{Collective-Excitations-2020,Twisted-bilayeriv-2021,Twisted-bilayerv-2021,
Lattice-collective-2021}.
The valley-wave modes have an extremely small excitation gap for the pristine TBG\cite{Lattice-collective-2021}, and
a sublattice potential difference increases their excitation energies\cite{Collective-Excitations-2020}.
For the pristine TBG,
low energy exciton states of a few meV also appear\cite{Lattice-collective-2021}.
We note that all the previous calculations
focused on one odd $\nu$ of -3 or 3 and adopted the active-band approximation that considers only the
excitations between flat bands\cite{Collective-Excitations-2020, Lattice-collective-2021}.
A full HF calculation of the excitation states at all odd $\nu$
may provide more excitation modes and can influence the excitation energy spectrums.
For TBG/BN with enhanced Chern insulator states at odd  $\nu$, previous studies have established that
BN induces not only the sublattice potential difference in graphene but also
spatially varying effective moir\'{e} potentials, and
the structural deformation due to the interlayer vdW interaction between BN and graphene
also strongly breaks the C$_{2z}$ symmetry of the single-particle Hamiltonian\cite{Symmetry-breaking-2020lin,Misalignment2021Lin}.
Moreover, the correlated band structure of TBG/BN changes
with the twist angle ($\theta'$) between TBG and BN\cite{Symmetry-breaking-2020lin,Band-structure-2020}.
Therefore, it is desirable to explore systematically the collective excitation modes at
all odd $\nu$ for all the possible commensurate configurations of TBG/BN.

Here, we demonstrate that the energies of the lowest intra-flavor exciton modes of TBG/BN
are much higher than those of TBG and reach about 20 meV, the inter-flavor
valley-wave modes have excitation energies higher than 2.5 meV which
is also much larger than that of TBG, while
the spin-wave modes all have zero excitation gap.
The excitation spectrums and gaps of TBG/BN at positive $\nu$ are rather different
from those at negative $\nu$, which contrasts with the
particle-hole symmetric excitation modes for positive and negative $\nu$ in TBG.
Full HF calculations indicate that the lowest exciton modes
that determine the optical properties of the Chern insulator states in TBG/BN are
generally the ones between the remote and flat-like bands, while the valley-wave modes have similar energies
as those in the active-band approximation. Moreover, the quantitative behavior of the excitation spectrum of TBG/BN
also varies with $\theta'$.

\section{HF bands and excitations in the active-band approximation}

For TBG with the magic twist angle of $\theta = 1.08^{\circ}$ aligned with BN,
we consider the $1 \times 1$ commensurate supercells of
TBG/BN at three twist angles $\theta'$ between BN and its adjacent graphene
layer of $1.64^{\circ}$, $0.54^{\circ}$, $-0.56^{\circ}$, as seen in Fig. 1(a),
and their structural parameters are detailed in the Supplemental Material (SM).
At the origin of the TBG/BN supercell, both the local stackings between
the graphene layers and between graphene and BN are taken to be the AA stacking.
The moir\'{e} superlattices of TBG/BN and the pristine TBG are fully
relaxed based on the continuum elastic theory to obtain their
stable atomic structures\cite{Symmetry-breaking-2020lin,Misalignment2021Lin}.

For the fully relaxed TBG/BN, an effective single-particle tight-binding model ($H^0$)
of the graphene layers can be built taking into account the relaxation effect and the full moir\'{e}
Hamiltonian induced by BN\cite{Symmetry-breaking-2020lin,Misalignment2021Lin}.
The parameters in $H^0$ and its expression in the planewave-like basis are detailed in the SM.
The single-particle flat bands around the Fermi level ($E_F$) in TBG are well separated by
the effective moir\'{e} potentials induced by BN, as shown in Fig. 1(b).
The C$_{2z}$ symmetry in the pristine TBG is broken in TBG/BN. In a rigid TBG/BN, the
effective Hamiltonian induced by BN lacks the C$_{2z}$ symmetry as reflected in the in-plane inversion asymmetric
terms of the moir\'{e} potentials. The structural relaxation of TBG/BN also leads to
the in-plane atomic deformation without the C$_{2z}$ symmetry.
The strength of the effective moir\'{e} potential by BN varies with $\theta'$,
giving rising to $\theta'$-dependent flat bands, as shown in Fig. S1(a)
of the SM. The widths of the flat bands are much larger at
$\theta'$ = $0.54^{\circ}$ and $-0.56^{\circ}$ than those at $1.64^{\circ}$,
while the valence and conduction bands are more separated at $\theta' = -0.56^{\circ}$.
The system at $\theta' = -0.56^{\circ}$ also has a much smaller energy
difference between the flat and remote bands due to the
wider flat bands and the larger gap at $E_F$.

Upon inclusion of the e-e interaction, TBG/BN and TBG become Chern insulators at odd $\nu$.
We employ the self-consistent HF (SCHF) method\cite{Nature2020xie,Symmetry-breaking-2020lin}
to obtain the mean-field ground states of
the systems at odd $\nu$, then the time dependent HF (TDHF)
approach\cite{Collective-Excitations-2020,Lattice-collective-2021} is adopted to
explore the collective excitations of TBG/BN and TBG based on the SCHF ground states as
detailed in the Appendix.
We first perform the HF calculations in the active-band approximation,
and the computationally expensive full HF calculations are then done for
further exploration of the collective excitations as presented in the next section.
For the active-space approximation, only the two central HF bands of each flavor are updated
during the SCHF iterations and they are only expanded in the basis
of the single-particle flat bands, and the lower remote bands are kept frozen but still
contribute to the mean-field Hartree and Fock operators of the active-band
Hamiltonian. In addition, the HF operators contributed by the isolated fixed and rotated
graphene layers with $E_F$ at CNP are subtracted from the HF Hamiltonian to
avoid double-counting of the e-e interaction.

The HF band structures of the Chern-insulator ground states at odd $\nu$ are exhibited in Fig. 1(c)
for TBG/BN with $\theta' = 0.54^{\circ}$ and the pristine TBG.
In TBG, sublattice polarization within one layer spontaneously occurs at odd $\nu$.
In the $\xi = +$ valley, the lower band has a Chern number of $+1$ and the higher band has a Chern number of $-1$.
The Chern numbers of the bands in the $\xi = -$ valley are just opposite
to those in the $\xi = +$ valley.
At each odd $\nu$, the ground states of TBG and TBG/BN are Chern insulators with the total Chern numbers of $\pm1$.
For each $\nu$, three of the four flavors have the same quantitative band properties,
such as the intra-flavor band gaps and the band widths, and one flavor has
different properties, which is taken to be the $(+,\uparrow)$ valley-spin flavor, as shown in Fig. 1(c).
At a flavor with the two bands both filled or empty, the two bands of TBG/BN are well separated, while those of the
pristine TBG have close energies around the $\bar{K}$ point. For TBG/BN at $\nu = -3$, the two empty bands
in the same flavor are separated by 17 meV.
When one flat band is filled and the other one is empty in a flavor,
the intra-flavor band gap ($\Delta_{intra}$) between them in TBG/BN is much larger than that in the pristine TBG.
Compared to $\Delta_{intra}$, the inter-favor band gap ($\Delta_{inter}$)
between the highest valence band in one flavor and the lowest conduction band in another flavor
generally has a smaller value, so the global band gap is just $\Delta_{inter}$.

The HF bands at $\nu = 3$ and $\nu = 1$ appear to be the particle-hole symmetric  correspondences of the bands
at $\nu = -3$ and $\nu = -1$, respectively, while
the band gaps can still be quite different between positive and negative $\nu$,
as shown in Figs. 1(c) and 2(b).
The $\theta'$ of TBG/BN influences the quantitative properties of the HF bands, as seen in Fig. S1(b).
At $\nu = \pm 3$, $\Delta_{inter}$ at $\theta' = 0.54^{\circ}$ is much larger
than those at other $\theta'$.
The $\Delta_{intra}$ at $\theta' = -0.56^{\circ}$ is the largest for $\nu = 3$.
In addition, when two bands in a flavor are both filled or empty,
they have a much larger energy difference at $\theta' = -0.56^{\circ}$.

We employ the TDHF method to obtain the collective excitation modes based on
the HF ground states at odd $\nu$.
We consider the collective modes with the momentum $\mb{q}$ expressed
as\cite{Collective-Excitations-2020, Lattice-collective-2021}
\begin{eqnarray}
|\Psi_{\{I\}}(\mb{q})\rangle =  \sum_{I,\mb{k}} u_{I,\mb{k}}(\mb{q}) f^{\dag}_{p_{I},\mb{k}+\mb{q}} f_{h_{I},\mb{k}} |0\rangle,
\end{eqnarray}
where $|0\rangle$ is the HF ground state, $I$ represents
an excitation process from the occupied band with index $h_{I}$
to the empty band with the index $p_{\mr{I}}$, and the operator $f$
annihilates an electron in the HF band states. A collective mode is characterized by
its set of excitation processes, which are labeled in Fig. 1(c) for the inter-flavor
spin-wave, valley-wave, and intra-flavor exciton modes.

For the pristine TBG, all the excitation spectrums exhibit approximate particle-hole
symmetry and are almost the same for all odd $\nu$, as shown in Fig. 2(a).
The spin-wave mode has a zero excitation gap, the valley-wave mode has an extremely small gap
of about 0.5 meV, and the exciton mode has a gap of about 5 meV.
Such finite gaps of the valley-wave and exciton modes are slightly larger than those predicted
for the TBG described by the Bistritzer-MacDonald model\cite{Lattice-collective-2021}, which can be attributed to the
in-plane structural deformation in the relaxed TBG. For the pristine TBG,
both the spin-wave and valley-wave excitations
have two low-energy collective modes in the gap at each $\mb{q}$.
In contrast, the excitation spectrum of TBG/BN at positive $\nu$ are rather different
from those at negative $\nu$, and those
with the same sign of $\nu$ are quite similar.
The lowest spin-wave mode at positive $\nu$ has a larger spectrum width than that
at negative $\nu$ but they all have zero excitation gap.
For the valley-wave, all the excitation energies are higher than 2.5 meV,
which is much larger than that of the pristine TBG. This is consistent with Ref.~[\onlinecite{Collective-Excitations-2020}] for TBG
with a sublattice potential difference.
The valley wave has a higher excitation gap ($\tilde{\Delta}_{inter}$) at positive $\nu$.
For both the spin wave and the valley wave, the lowest modes become much more apart than those in TBG.
The lowest exciton modes of TBG/BN have much higher energies than those of TBG.
The exciton gap ($\tilde{\Delta}_{intra}$) decreases with $\nu$ from -3 to 3,
with the gap still reaching about 16 meV at $\nu = 3$.

In comparison to the HF band gaps, the $\tilde{\Delta}_{inter}$ of the valley wave modes
are much smaller than the ${\Delta}_{inter}$ for both TBG and
TBG/BN, while the $\tilde{\Delta}_{intra}$ of the exciton modes of TBG/BN
reaches about half of the ${\Delta}_{intra}$, which is a significant contrast to
the much smaller $\tilde{\Delta}_{intra}/{\Delta}_{intra}$ for TBG, as shown in Fig 2(b).
For the exciton modes, the two-body exciton wavefunction as a function of
the electron ($\mb{r}_e$) and hole ($\mb{r}_h$) positions can be calculated as
\begin{eqnarray}
\Psi(\mb{r}_e, \mb{r}_h) = \frac{1}{N_{\mb{k}}} \sum_{\mb{k}} u_{\mb{k}} \psi_{p\mb{k}}(\mb{r}_e) \psi^*_{h\mb{k}}(\mb{r}_h),
\end{eqnarray}
where $\psi_{p\mb{k}}$ and $\psi_{h\mb{k}}$ are the HF conduction and
valence band states corresponding to the exciton excitation.
Figure 3 exhibits the wavefunction of the lowest exciton mode at $\mb{q} = \mb{0}$
with $\mb{r}_h$ at the origin of a supercell where the bilayer graphene is locally AA-stacked.
The particle and hole are strongly bound at all the odd $\nu$ for the pristine TBG with the
particle localized around the origin.
The spatial map of the exciton wavefunctions in TBG/BN spread a much larger range with
the particle mainly distributed around the nonzero smallest superlattice vectors.
Unlike TBG, TBG/BN at $\nu = 3$ has a quite different exciton wavefunction from that
at $\nu = -3$ with the wavefunction at $\nu = 3$ less spatially localized.

The quantitative behavior of the excitation spectrum varies with $\theta'$, as seen in Fig. 4.
The systems with the negative $\theta'$ of $-0.56^{\circ}$ tend to have a smaller
valley-wave excitation energy, while their exciton energies are much higher than those at positive
$\theta'$ for the positive $\nu$. For the valley-wave modes, the $\tilde{\Delta}_{inter}$
at the two positive $\theta'$ have similar values for the negative $\nu$, while
they differ by about 1 meV for the positive $\nu$. The exciton energy at $\theta' = 0.54^{\circ}$
is higher than that at $\theta' = 1.64^{\circ}$ for the negative $\nu$ but has
similar values for the positive $\nu$.

\section{Full HF bands and excitations}

Since the remote bands are frozen in the active-band approximation of the SCHF calculations,
the excitation processes between the remote and flat bands have been ignored, and the
quantitative properties of the flat bands can be modified when the remote bands are allowed to
be updated in the SCHF calculations. Full SCHF calculations have also been performed to
obtain the full HF bands of TBG/BN, and the excitation spectrums are computed by considering
the excitation processes between the five highest valence HF bands and the five lowest conduction HF bands.
It is noted that the convergent spin-wave spectrum requires the possible excitation processes
between all the HF bands, which are beyond our calculation capability, so
only the inter-flavor valley-wave and the intra-flavor exciton modes are considered
based on the full SCHF ground states.

To compare the active-band approximation with the full SCHF description of the central HF bands,
the projection of each HF band state on the single-particle flat bands is
computed as $\sum_{m} |\langle \psi_m^0(\sigma,\mb{k}) | \psi_i(\sigma,\mb{k}) \rangle|^2$ with $|\psi_m^0(\sigma,\mb{k})\rangle$
representing the two single-particle flat-band states of flavor $\sigma$ and $|\psi_i \rangle$
a HF band state. We find that at a k-point rather away from the $\bar{\Gamma}$ point, only two low-energy
HF band states of a flavor are mainly contributed by the single-particle flat-band states,
as shown in Fig. 5 for TBG/BN with $\theta' = 0.54^{\circ}$.
These HF bands are termed as flat-like bands to distinguish them from the single-particle flat bands.
In contrast, several other HF bands near the $\bar{\Gamma}$ point can have substantial contribution from
the flat-band states, especially for the flavor with one flat-like band occupied and the other flat-like band empty.
In particular, the flat-band contribution becomes very small for some low-energy HF states at $\bar{\Gamma}$.
When the flat-like bands of a flavor are both occupied or empty, they are generally well separated
from the remote bands, and the intra-flavor gap around $E_F$ between the remote and flat-like bands
is denoted by $\Delta'_{intra}$.
The flat-like bands become entangled with the remote bands when $E_F$ lies between them, and
the intra-flavor gap between these flat-like bands is denoted by $\Delta_{intra}$.
Similar to the active-band approximation, the inter-flavor gap $\Delta_{inter}$ is also between
the flat-like bands. For the full HF bands, the global gap among all flavors is just
$\Delta'_{intra}$. $\Delta_{intra}$ has large and similar values for all the negative and
positive $\nu$, which is similar to the active-band approximation.
However, the systems at positive $\nu$ have much smaller $\Delta'_{intra}$ and $\Delta_{inter}$
than those at negative $\nu$, which indicates the strong breaking of the particle-hole symmetry
for the full HF band structures. At each $\nu$, there are also three flavors with the same quantitative
band properties, as seen in Fig. 5.

We consider the inter-flavor valley-wave excitation modes corresponding to $\Delta_{inter}$,
and the intra-flavor exciton modes corresponding to $\Delta_{intra}$ and
$\Delta'_{intra}$, based on the full HF ground states. The excitation spectrum and the
excitation gaps of these modes are displayed in Fig. 6 for TBG/BN with $\theta' = 0.54^{\circ}$.
The valley-wave excitation gap $\tilde{\Delta}_{inter}$ becomes
slightly higher than that obtained from the active-band approximation and reaches about 3 meV,
but is still rather small compared with $\Delta_{inter}$. The excitation gaps
$\tilde{\Delta}_{intra}$ of the exciton modes between the flat-like bands have similar values as
those from the active-band approximation and are below half of $\Delta_{intra}$. In contrast,
The gaps ($\tilde{\Delta}'_{intra}$) of the exciton modes between the flat-like bands and the remote bands
are just slightly smaller than ${\Delta}'_{intra}$. This indicates that the exciton modes between the
flat-like bands and the remote bands are composed of weak-bound particle-hole pairs, while
strong binding of the particle-hole pairs occurs in the exciton modes between the flat-like bands.
At $\nu = -3, 1, 3$, $\tilde{\Delta}_{intra}$ is higher than $\tilde{\Delta}'_{intra}$ and
even the gap $\Delta'_{intra}$. Only at $\nu = 1$, $\tilde{\Delta}_{intra}$ has a lower value
than $\tilde{\Delta}'_{intra}$. In addition, the excitation energies of the lowest modes
for the exciton modes between the flat-like bands and the remote bands are much more dispersive
as a function of $\mb{q}$ than those of the valley-wave modes and the
exciton modes between the flat-like bands, as shown in Fig. 6(b).

The optical properties of the Chern insulators are determined by the intra-flavor exciton modes,
and the optical conductivity within the TDHF method is given by\cite{Lattice-collective-2021}
\begin{eqnarray}
\mr{Re}\sigma_{xx}  &=& \frac{\gamma}{\hbar\omega N_{\mb{k}} \Omega_0} \sum_{i} \frac{1}{(\hbar\omega - \hbar\omega_i)^2 + \gamma^2} \nonumber \\
&&\sum_{I\mb{k}, I'\mb{k}'} J_{x,I\mb{k}}^* u_{i, I\mb{k}} u_{i, I'\mb{k}'}^* J_{x,I'\mb{k}'}
\end{eqnarray}
where $\omega$ is the frequency of the incident light, $\hbar\omega_i$ is the energy of an
exciton mode labeled by $i$, $u_{i}$ is the state vector of the exciton mode,
$J_{x,I\mb{k}} = \langle \psi_{p_I\mb{k}}|-e/\hbar \partial H_{k}/\partial k_x|\psi_{h_I\mb{k}} \rangle$
is the element of the current density operator between the empty and occupied
states of the excitation process $I$, $\gamma$ is a small energy for broadening
of the excitation energy, $\Omega_0$ is the area of the moire supercell, and
$N_{\mb{k}}$ is the number of k-points. So at $\omega = \omega_i$, the contribution
of the exciton mode $i$ to $\sigma_{xx}$ is proportional to
$\sigma_i \equiv \hbar^2/(e^2 N_{\mb{k}}) \sum_{I\mb{k}, I'\mb{k}'} J_{x,I\mb{k}}^* u_{i, I\mb{k}} u_{i, I'\mb{k}'}^* J_{x,I'\mb{k}'}$.
We find that the $\sigma_i$ of the lowest exciton mode between the remote and flat-like bands at $\nu = -3$
reaches 0.102 eV {\AA}$^2$ and is even much larger than that
of the lowest exciton mode between the flat-like bands, which is just 0.022 eV {\AA}$^2$.
Therefore, the lowest-frequency optical properties associated with the intra-flavor excitations
are mainly determined by the exciton modes between the remote bands and the flat-like bands at $\nu = -3, 1, 3$,
while they are mainly contributed by the exciton mode between the flat-like bands at $\nu = -1$.

At the other two $\theta'$ of $1.64^{\circ}$ and $-0.56^{\circ}$, $\Delta'_{intra}$ from the full SCHF calculations
can become larger than $\Delta_{inter}$, but are all much smaller than
$\Delta_{intra}$, as seen in Figs. S2 and S3 of the SM. For $\theta' = -0.56^{\circ}$, the system at $\nu = -3$ becomes
metallic with the highest occupied band of the $(+,\uparrow)$ flavor slightly overlapping with
the lowest empty bands of other flavors. The systems at $\theta' = 1.64^{\circ}$ generally have
smaller $\Delta'_{intra}$ than those at other $\theta'$.
For the exciton modes, the excitation gaps $\tilde{\Delta}'_{intra}$ are also much
smaller than $\tilde{\Delta}_{intra}$, and the systems with $\theta' = -0.56^{\circ}$
have the largest $\tilde{\Delta}_{intra}$, as seen in Fig. S3 of SM.
In addition, $\tilde{\Delta}'_{intra}$ can even become larger than the indirect gap $\Delta'_{intra}$
for some systems with $\theta'$ of $1.64^{\circ}$ and $-0.56^{\circ}$.
The $\tilde{\Delta}_{inter}$ for the valley-wave modes all have similar values of about 3 meV.

\section{Summary and Conclusions}

In the $1 \times 1$ commensurate supercells of TBG/BN,
the single-particle flat bands around $E_F$ are gaped due to
the broken C$_{2z}$ symmetry, and the SCHF ground states at
odd $\nu$ are the Chern insulators
with flavor-polarized HF bands.
In the active-band approximation, the two active HF bands in
the same flavor are well separated in TBG/BN when they are both
filled or empty, and the intra-flavor gap $\Delta_{intra}$
in TBG/BN is much larger than that in the pristine TBG.
The energy spectrums of the collective excitation modes for the
Chern insulator states are obtained with the TDHF method.
The spin-wave modes in both TBG/BN and TBG have a zero excitation gap,
while the gaps of the valley-wave and exciton modes in TBG/BN are
much larger than those in TBG. The excitation gap $\tilde{\Delta}_{inter}$ and
$\tilde{\Delta}_{intra}$ in TBG/BN reach about 2.5 meV and 20 meV,
respectively, with $\tilde{\Delta}_{intra}$ almost a half of the
intra-flavor band gap ${\Delta}_{intra}$.
In contrast to TBG with almost particle-hole symmetric excitation modes
for positive and negative $\nu$,
the excitation spectrums and gaps of TBG/BN at positive $\nu$
are rather different from those at negative $\nu$.
The exciton wavefunctions in TBG are also much more spatially localized than those in TBG/BN.
Full SCHF calculations show that more HF bands besides
the two central bands can
have rather large contribution
from the single-particle flat-band states in TBG/BN, and
the intra-flavor gap $\Delta_{intra}$ between the flat-like bands
is much larger than the $\Delta'_{intra}$ between the remote and flat-like bands.
The excitation gap $\tilde{\Delta}'_{intra}$ of the exciton modes
between the remote and flat-like bands
is just slightly smaller than $\Delta'_{intra}$,
but is generally lower than the $\tilde{\Delta}_{intra}$
between the flat-like bands, so
the optical properties of the Chern insulator states are mainly determined by the exciton modes between the
remote and flat-like bands.
The valley-wave modes from full HF calculations have similar energies
as those in the active-band approximation.
In addition, the quantitative behavior of the excitation spectrums
varies with $\theta'$ of TBG/BN.

\label{Acknowledgments}
\begin{acknowledgments}
We gratefully acknowledge valuable discussions with D. Tom\'anek, Y. Yin, and X. Xiong.
This research was supported by
the National Natural Science Foundation of China (Grants No. 11974312 and No. 92270104)
and the Open Research Fund of CNMGE Platform \& NSCC-TJ.
\end{acknowledgments}


\begin{thebibliography}{64}%
\makeatletter
\providecommand \@ifxundefined [1]{%
 \@ifx{#1\undefined}
}%
\providecommand \@ifnum [1]{%
 \ifnum #1\expandafter \@firstoftwo
 \else \expandafter \@secondoftwo
 \fi
}%
\providecommand \@ifx [1]{%
 \ifx #1\expandafter \@firstoftwo
 \else \expandafter \@secondoftwo
 \fi
}%
\providecommand \natexlab [1]{#1}%
\providecommand \enquote  [1]{``#1''}%
\providecommand \bibnamefont  [1]{#1}%
\providecommand \bibfnamefont [1]{#1}%
\providecommand \citenamefont [1]{#1}%
\providecommand \href@noop [0]{\@secondoftwo}%
\providecommand \href [0]{\begingroup \@sanitize@url \@href}%
\providecommand \@href[1]{\@@startlink{#1}\@@href}%
\providecommand \@@href[1]{\endgroup#1\@@endlink}%
\providecommand \@sanitize@url [0]{\catcode `\\12\catcode `\$12\catcode
  `\&12\catcode `\#12\catcode `\^12\catcode `\_12\catcode `\%12\relax}%
\providecommand \@@startlink[1]{}%
\providecommand \@@endlink[0]{}%
\providecommand \url  [0]{\begingroup\@sanitize@url \@url }%
\providecommand \@url [1]{\endgroup\@href {#1}{\urlprefix }}%
\providecommand \urlprefix  [0]{URL }%
\providecommand \Eprint [0]{\href }%
\providecommand \doibase [0]{http://dx.doi.org/}%
\providecommand \selectlanguage [0]{\@gobble}%
\providecommand \bibinfo  [0]{\@secondoftwo}%
\providecommand \bibfield  [0]{\@secondoftwo}%
\providecommand \translation [1]{[#1]}%
\providecommand \BibitemOpen [0]{}%
\providecommand \bibitemStop [0]{}%
\providecommand \bibitemNoStop [0]{.\EOS\space}%
\providecommand \EOS [0]{\spacefactor3000\relax}%
\providecommand \BibitemShut  [1]{\csname bibitem#1\endcsname}%
\let\auto@bib@innerbib\@empty
\bibitem [{\citenamefont {Bistritzer}\ and\ \citenamefont
  {MacDonald}(2011)}]{Bistritzer12233}%
  \BibitemOpen
  \bibfield  {author} {\bibinfo {author} {\bibfnamefont {R.}~\bibnamefont
  {Bistritzer}}\ and\ \bibinfo {author} {\bibfnamefont {A.~H.}\ \bibnamefont
  {MacDonald}},\ }\bibfield  {title} {\enquote {\bibinfo {title} {{M}oir{\'e}
  bands in twisted double-layer graphene},}\ }\href {\doibase
  10.1073/pnas.1108174108} {\bibfield  {journal} {\bibinfo  {journal} {Proc.
  Natl. Acad. Sci. U.S.A.}\ }\textbf {\bibinfo {volume} {108}},\ \bibinfo
  {pages} {12233} (\bibinfo {year} {2011})}\BibitemShut {NoStop}%
\bibitem [{\citenamefont {Su\'arez~Morell}\ \emph {et~al.}(2010)\citenamefont
  {Su\'arez~Morell}, \citenamefont {Correa}, \citenamefont {Vargas},
  \citenamefont {Pacheco},\ and\ \citenamefont {Barticevic}}]{Morell2010}%
  \BibitemOpen
  \bibfield  {author} {\bibinfo {author} {\bibfnamefont {E.}~\bibnamefont
  {Su\'arez~Morell}}, \bibinfo {author} {\bibfnamefont {J.~D.}\ \bibnamefont
  {Correa}}, \bibinfo {author} {\bibfnamefont {P.}~\bibnamefont {Vargas}},
  \bibinfo {author} {\bibfnamefont {M.}~\bibnamefont {Pacheco}}, \ and\
  \bibinfo {author} {\bibfnamefont {Z.}~\bibnamefont {Barticevic}},\ }\bibfield
   {title} {\enquote {\bibinfo {title} {Flat bands in slightly twisted bilayer
  graphene: Tight-binding calculations},}\ }\href {\doibase
  10.1103/PhysRevB.82.121407} {\bibfield  {journal} {\bibinfo  {journal} {Phys.
  Rev. B}\ }\textbf {\bibinfo {volume} {82}},\ \bibinfo {pages} {121407}
  (\bibinfo {year} {2010})}\BibitemShut {NoStop}%
\bibitem [{\citenamefont {Lopes~dos Santos}\ \emph {et~al.}(2012)\citenamefont
  {Lopes~dos Santos}, \citenamefont {Peres},\ and\ \citenamefont
  {Castro~Neto}}]{LopesdosSantos2012}%
  \BibitemOpen
  \bibfield  {author} {\bibinfo {author} {\bibfnamefont {J.~M.~B.}\
  \bibnamefont {Lopes~dos Santos}}, \bibinfo {author} {\bibfnamefont
  {N.~M.~R.}\ \bibnamefont {Peres}}, \ and\ \bibinfo {author} {\bibfnamefont
  {A.~H.}\ \bibnamefont {Castro~Neto}},\ }\bibfield  {title} {\enquote
  {\bibinfo {title} {Continuum model of the twisted graphene bilayer},}\ }\href
  {\doibase 10.1103/PhysRevB.86.155449} {\bibfield  {journal} {\bibinfo
  {journal} {Phys. Rev. B}\ }\textbf {\bibinfo {volume} {86}},\ \bibinfo
  {pages} {155449} (\bibinfo {year} {2012})}\BibitemShut {NoStop}%
\bibitem [{\citenamefont {Fang}\ and\ \citenamefont
  {Kaxiras}(2016)}]{Fang2016}%
  \BibitemOpen
  \bibfield  {author} {\bibinfo {author} {\bibfnamefont {S.}~\bibnamefont
  {Fang}}\ and\ \bibinfo {author} {\bibfnamefont {E.}~\bibnamefont {Kaxiras}},\
  }\bibfield  {title} {\enquote {\bibinfo {title} {Electronic structure theory
  of weakly interacting bilayers},}\ }\href {\doibase
  10.1103/PhysRevB.93.235153} {\bibfield  {journal} {\bibinfo  {journal} {Phys.
  Rev. B}\ }\textbf {\bibinfo {volume} {93}},\ \bibinfo {pages} {235153}
  (\bibinfo {year} {2016})}\BibitemShut {NoStop}%
\bibitem [{\citenamefont {Carr}\ \emph {et~al.}(2017)\citenamefont {Carr},
  \citenamefont {Massatt}, \citenamefont {Fang}, \citenamefont {Cazeaux},
  \citenamefont {Luskin},\ and\ \citenamefont {Kaxiras}}]{Carr2017}%
  \BibitemOpen
  \bibfield  {author} {\bibinfo {author} {\bibfnamefont {S.}~\bibnamefont
  {Carr}}, \bibinfo {author} {\bibfnamefont {D.}~\bibnamefont {Massatt}},
  \bibinfo {author} {\bibfnamefont {S.}~\bibnamefont {Fang}}, \bibinfo {author}
  {\bibfnamefont {P.}~\bibnamefont {Cazeaux}}, \bibinfo {author} {\bibfnamefont
  {M.}~\bibnamefont {Luskin}}, \ and\ \bibinfo {author} {\bibfnamefont
  {E.}~\bibnamefont {Kaxiras}},\ }\bibfield  {title} {\enquote {\bibinfo
  {title} {{T}wistronics: {M}anipulating the electronic properties of
  two-dimensional layered structures through their twist angle},}\ }\href
  {\doibase 10.1103/PhysRevB.95.075420} {\bibfield  {journal} {\bibinfo
  {journal} {Phys. Rev. B}\ }\textbf {\bibinfo {volume} {95}},\ \bibinfo
  {pages} {075420} (\bibinfo {year} {2017})}\BibitemShut {NoStop}%
\bibitem [{\citenamefont {Tarnopolsky}\ \emph {et~al.}(2019)\citenamefont
  {Tarnopolsky}, \citenamefont {Kruchkov},\ and\ \citenamefont
  {Vishwanath}}]{OriginPhysRevLett.122.106405}%
  \BibitemOpen
  \bibfield  {author} {\bibinfo {author} {\bibfnamefont {G.}~\bibnamefont
  {Tarnopolsky}}, \bibinfo {author} {\bibfnamefont {A.~Jura}\ \bibnamefont
  {Kruchkov}}, \ and\ \bibinfo {author} {\bibfnamefont {A.}~\bibnamefont
  {Vishwanath}},\ }\bibfield  {title} {\enquote {\bibinfo {title} {Origin of
  {M}agic {A}ngles in {T}wisted {B}ilayer {G}raphene},}\ }\href {\doibase
  10.1103/PhysRevLett.122.106405} {\bibfield  {journal} {\bibinfo  {journal}
  {Phys. Rev. Lett.}\ }\textbf {\bibinfo {volume} {122}},\ \bibinfo {pages}
  {106405} (\bibinfo {year} {2019})}\BibitemShut {NoStop}%
\bibitem [{\citenamefont {Ren}\ \emph {et~al.}(2021)\citenamefont {Ren},
  \citenamefont {Gao}, \citenamefont {MacDonald},\ and\ \citenamefont
  {Niu}}]{WKB-Estimate-2021}%
  \BibitemOpen
  \bibfield  {author} {\bibinfo {author} {\bibfnamefont {Y.}~\bibnamefont
  {Ren}}, \bibinfo {author} {\bibfnamefont {Q.}~\bibnamefont {Gao}}, \bibinfo
  {author} {\bibfnamefont {A.~H.}\ \bibnamefont {MacDonald}}, \ and\ \bibinfo
  {author} {\bibfnamefont {Q.}~\bibnamefont {Niu}},\ }\bibfield  {title}
  {\enquote {\bibinfo {title} {{W}{K}{B} {E}stimate of {B}ilayer {G}raphene's
  {M}agic {T}wist {A}ngles},}\ }\href {\doibase 10.1103/PhysRevLett.126.016404}
  {\bibfield  {journal} {\bibinfo  {journal} {Phys. Rev. Lett.}\ }\textbf
  {\bibinfo {volume} {126}},\ \bibinfo {pages} {016404} (\bibinfo {year}
  {2021})}\BibitemShut {NoStop}%
\bibitem [{\citenamefont {Cao}\ \emph {et~al.}(2018{\natexlab{a}})\citenamefont
  {Cao}, \citenamefont {Fatemi}, \citenamefont {Demir}, \citenamefont {Fang},
  \citenamefont {Tomarken}, \citenamefont {Luo}, \citenamefont
  {Sanchez-Yamagishi}, \citenamefont {Watanabe}, \citenamefont {Taniguchi},
  \citenamefont {Kaxiras}, \citenamefont {Ashoori},\ and\ \citenamefont
  {Jarillo-Herrero}}]{Cao2018}%
  \BibitemOpen
  \bibfield  {author} {\bibinfo {author} {\bibfnamefont {Y.}~\bibnamefont
  {Cao}}, \bibinfo {author} {\bibfnamefont {V.}~\bibnamefont {Fatemi}},
  \bibinfo {author} {\bibfnamefont {A.}~\bibnamefont {Demir}}, \bibinfo
  {author} {\bibfnamefont {S.}~\bibnamefont {Fang}}, \bibinfo {author}
  {\bibfnamefont {S.~L.}\ \bibnamefont {Tomarken}}, \bibinfo {author}
  {\bibfnamefont {J.~Y.}\ \bibnamefont {Luo}}, \bibinfo {author} {\bibfnamefont
  {J.~D.}\ \bibnamefont {Sanchez-Yamagishi}}, \bibinfo {author} {\bibfnamefont
  {K.}~\bibnamefont {Watanabe}}, \bibinfo {author} {\bibfnamefont
  {T.}~\bibnamefont {Taniguchi}}, \bibinfo {author} {\bibfnamefont
  {E.}~\bibnamefont {Kaxiras}}, \bibinfo {author} {\bibfnamefont {R.~C.}\
  \bibnamefont {Ashoori}}, \ and\ \bibinfo {author} {\bibfnamefont
  {P.}~\bibnamefont {Jarillo-Herrero}},\ }\bibfield  {title} {\enquote
  {\bibinfo {title} {Correlated insulator behaviour at half-filling in
  magic-angle graphene superlattices},}\ }\href {\doibase 10.1038/nature26154}
  {\bibfield  {journal} {\bibinfo  {journal} {Nature}\ }\textbf {\bibinfo
  {volume} {556}},\ \bibinfo {pages} {80} (\bibinfo {year}
  {2018}{\natexlab{a}})}\BibitemShut {NoStop}%
\bibitem [{\citenamefont {Cao}\ \emph {et~al.}(2018{\natexlab{b}})\citenamefont
  {Cao}, \citenamefont {Fatemi}, \citenamefont {Fang}, \citenamefont
  {Watanabe}, \citenamefont {Taniguchi}, \citenamefont {Kaxiras},\ and\
  \citenamefont {Jarillo-Herrero}}]{cao2018unconventional}%
  \BibitemOpen
  \bibfield  {author} {\bibinfo {author} {\bibfnamefont {Y.}~\bibnamefont
  {Cao}}, \bibinfo {author} {\bibfnamefont {V.}~\bibnamefont {Fatemi}},
  \bibinfo {author} {\bibfnamefont {S.}~\bibnamefont {Fang}}, \bibinfo {author}
  {\bibfnamefont {K.}~\bibnamefont {Watanabe}}, \bibinfo {author}
  {\bibfnamefont {T.}~\bibnamefont {Taniguchi}}, \bibinfo {author}
  {\bibfnamefont {E.}~\bibnamefont {Kaxiras}}, \ and\ \bibinfo {author}
  {\bibfnamefont {P.}~\bibnamefont {Jarillo-Herrero}},\ }\bibfield  {title}
  {\enquote {\bibinfo {title} {Unconventional superconductivity in magic-angle
  graphene superlattices},}\ }\href {http://dx.doi.org/10.1038/nature26160}
  {\bibfield  {journal} {\bibinfo  {journal} {Nature}\ }\textbf {\bibinfo
  {volume} {556}},\ \bibinfo {pages} {43} (\bibinfo {year}
  {2018}{\natexlab{b}})}\BibitemShut {NoStop}%
\bibitem [{\citenamefont {Kim}\ \emph {et~al.}(2017)\citenamefont {Kim},
  \citenamefont {DaSilva}, \citenamefont {Huang}, \citenamefont {Fallahazad},
  \citenamefont {Larentis}, \citenamefont {Taniguchi}, \citenamefont
  {Watanabe}, \citenamefont {LeRoy}, \citenamefont {MacDonald},\ and\
  \citenamefont {Tutuc}}]{Kim3364}%
  \BibitemOpen
  \bibfield  {author} {\bibinfo {author} {\bibfnamefont {K.}~\bibnamefont
  {Kim}}, \bibinfo {author} {\bibfnamefont {A.}~\bibnamefont {DaSilva}},
  \bibinfo {author} {\bibfnamefont {S.}~\bibnamefont {Huang}}, \bibinfo
  {author} {\bibfnamefont {B.}~\bibnamefont {Fallahazad}}, \bibinfo {author}
  {\bibfnamefont {S.}~\bibnamefont {Larentis}}, \bibinfo {author}
  {\bibfnamefont {T.}~\bibnamefont {Taniguchi}}, \bibinfo {author}
  {\bibfnamefont {K.}~\bibnamefont {Watanabe}}, \bibinfo {author}
  {\bibfnamefont {B.~J.}\ \bibnamefont {LeRoy}}, \bibinfo {author}
  {\bibfnamefont {A.~H.}\ \bibnamefont {MacDonald}}, \ and\ \bibinfo {author}
  {\bibfnamefont {E.}~\bibnamefont {Tutuc}},\ }\bibfield  {title} {\enquote
  {\bibinfo {title} {Tunable {M}oir{\'e} bands and strong correlations in
  small-twist-angle bilayer graphene},}\ }\href {\doibase
  10.1073/pnas.1620140114} {\bibfield  {journal} {\bibinfo  {journal} {Proc.
  Natl. Acad. Sci. U.S.A}\ }\textbf {\bibinfo {volume} {114}},\ \bibinfo
  {pages} {3364} (\bibinfo {year} {2017})}\BibitemShut {NoStop}%
\bibitem [{\citenamefont {Lu}\ \emph {et~al.}(2019)\citenamefont {Lu},
  \citenamefont {Stepanov}, \citenamefont {Yang}, \citenamefont {Xie},
  \citenamefont {Aamir}, \citenamefont {Das}, \citenamefont {Urgell},
  \citenamefont {Watanabe}, \citenamefont {Taniguchi}, \citenamefont {Zhang},
  \citenamefont {Bachtold}, \citenamefont {MacDonald},\ and\ \citenamefont
  {Efetov}}]{lu2019superconductors}%
  \BibitemOpen
  \bibfield  {author} {\bibinfo {author} {\bibfnamefont {X.}~\bibnamefont
  {Lu}}, \bibinfo {author} {\bibfnamefont {P.}~\bibnamefont {Stepanov}},
  \bibinfo {author} {\bibfnamefont {W.}~\bibnamefont {Yang}}, \bibinfo {author}
  {\bibfnamefont {M.}~\bibnamefont {Xie}}, \bibinfo {author} {\bibfnamefont
  {M.~A.}\ \bibnamefont {Aamir}}, \bibinfo {author} {\bibfnamefont
  {I.}~\bibnamefont {Das}}, \bibinfo {author} {\bibfnamefont {C.}~\bibnamefont
  {Urgell}}, \bibinfo {author} {\bibfnamefont {K.}~\bibnamefont {Watanabe}},
  \bibinfo {author} {\bibfnamefont {T.}~\bibnamefont {Taniguchi}}, \bibinfo
  {author} {\bibfnamefont {G.}~\bibnamefont {Zhang}}, \bibinfo {author}
  {\bibfnamefont {A.}~\bibnamefont {Bachtold}}, \bibinfo {author}
  {\bibfnamefont {A.~H.}\ \bibnamefont {MacDonald}}, \ and\ \bibinfo {author}
  {\bibfnamefont {D.~K.}\ \bibnamefont {Efetov}},\ }\bibfield  {title}
  {\enquote {\bibinfo {title} {Superconductors, orbital magnets, and correlated
  states in magic angle bilayer graphene},}\ }\href {\doibase
  10.1038/s41586-019-1695-0} {\bibfield  {journal} {\bibinfo  {journal}
  {Nature}\ }\textbf {\bibinfo {volume} {574}},\ \bibinfo {pages} {653}
  (\bibinfo {year} {2019})}\BibitemShut {NoStop}%
\bibitem [{\citenamefont {Xie}\ \emph {et~al.}(2019)\citenamefont {Xie},
  \citenamefont {Lian}, \citenamefont {J{\"a}ck}, \citenamefont {Liu},
  \citenamefont {Chiu}, \citenamefont {Watanabe}, \citenamefont {Taniguchi},
  \citenamefont {Bernevig},\ and\ \citenamefont
  {Yazdani}}]{xie2019spectroscopic}%
  \BibitemOpen
  \bibfield  {author} {\bibinfo {author} {\bibfnamefont {Y.}~\bibnamefont
  {Xie}}, \bibinfo {author} {\bibfnamefont {B.}~\bibnamefont {Lian}}, \bibinfo
  {author} {\bibfnamefont {B.}~\bibnamefont {J{\"a}ck}}, \bibinfo {author}
  {\bibfnamefont {X.}~\bibnamefont {Liu}}, \bibinfo {author} {\bibfnamefont
  {C.-L.}\ \bibnamefont {Chiu}}, \bibinfo {author} {\bibfnamefont
  {K.}~\bibnamefont {Watanabe}}, \bibinfo {author} {\bibfnamefont
  {T.}~\bibnamefont {Taniguchi}}, \bibinfo {author} {\bibfnamefont {B.~A.}\
  \bibnamefont {Bernevig}}, \ and\ \bibinfo {author} {\bibfnamefont
  {A.}~\bibnamefont {Yazdani}},\ }\bibfield  {title} {\enquote {\bibinfo
  {title} {Spectroscopic signatures of many-body correlations in magic-angle
  twisted bilayer graphene},}\ }\href {\doibase 10.1038/s41586-019-1422-x}
  {\bibfield  {journal} {\bibinfo  {journal} {Nature}\ }\textbf {\bibinfo
  {volume} {572}},\ \bibinfo {pages} {101} (\bibinfo {year}
  {2019})}\BibitemShut {NoStop}%
\bibitem [{\citenamefont {Kerelsky}\ \emph {et~al.}(2019)\citenamefont
  {Kerelsky}, \citenamefont {McGilly}, \citenamefont {Kennes}, \citenamefont
  {Xian}, \citenamefont {Yankowitz}, \citenamefont {Chen}, \citenamefont
  {Watanabe}, \citenamefont {Taniguchi}, \citenamefont {Hone}, \citenamefont
  {Dean}, \citenamefont {Rubio},\ and\ \citenamefont
  {Pasupathy}}]{Maximized-electron-2019}%
  \BibitemOpen
  \bibfield  {author} {\bibinfo {author} {\bibfnamefont {A.}~\bibnamefont
  {Kerelsky}}, \bibinfo {author} {\bibfnamefont {L.~J.}\ \bibnamefont
  {McGilly}}, \bibinfo {author} {\bibfnamefont {D.~M.}\ \bibnamefont {Kennes}},
  \bibinfo {author} {\bibfnamefont {L.}~\bibnamefont {Xian}}, \bibinfo {author}
  {\bibfnamefont {M.}~\bibnamefont {Yankowitz}}, \bibinfo {author}
  {\bibfnamefont {S.}~\bibnamefont {Chen}}, \bibinfo {author} {\bibfnamefont
  {K.}~\bibnamefont {Watanabe}}, \bibinfo {author} {\bibfnamefont
  {T.}~\bibnamefont {Taniguchi}}, \bibinfo {author} {\bibfnamefont
  {J.}~\bibnamefont {Hone}}, \bibinfo {author} {\bibfnamefont {C.}~\bibnamefont
  {Dean}}, \bibinfo {author} {\bibfnamefont {A.}~\bibnamefont {Rubio}}, \ and\
  \bibinfo {author} {\bibfnamefont {A.~N.}\ \bibnamefont {Pasupathy}},\
  }\bibfield  {title} {\enquote {\bibinfo {title} {Maximized electron
  interactions at the magic angle in twisted bilayer graphene},}\ }\href
  {\doibase 10.1038/s41586-019-1431-9} {\bibfield  {journal} {\bibinfo
  {journal} {Nature}\ }\textbf {\bibinfo {volume} {572}},\ \bibinfo {pages}
  {95} (\bibinfo {year} {2019})}\BibitemShut {NoStop}%
\bibitem [{\citenamefont {Jiang}\ \emph {et~al.}(2019)\citenamefont {Jiang},
  \citenamefont {Lai}, \citenamefont {Watanabe}, \citenamefont {Taniguchi},
  \citenamefont {Haule}, \citenamefont {Mao},\ and\ \citenamefont
  {Andrei}}]{Charge-order-2019}%
  \BibitemOpen
  \bibfield  {author} {\bibinfo {author} {\bibfnamefont {Y.}~\bibnamefont
  {Jiang}}, \bibinfo {author} {\bibfnamefont {X.}~\bibnamefont {Lai}}, \bibinfo
  {author} {\bibfnamefont {K.}~\bibnamefont {Watanabe}}, \bibinfo {author}
  {\bibfnamefont {T.}~\bibnamefont {Taniguchi}}, \bibinfo {author}
  {\bibfnamefont {K.}~\bibnamefont {Haule}}, \bibinfo {author} {\bibfnamefont
  {J.}~\bibnamefont {Mao}}, \ and\ \bibinfo {author} {\bibfnamefont {E.~Y.}\
  \bibnamefont {Andrei}},\ }\bibfield  {title} {\enquote {\bibinfo {title}
  {Charge order and broken rotational symmetry in magic-angle twisted bilayer
  graphene},}\ }\href {\doibase 10.1038/s41586-019-1460-4} {\bibfield
  {journal} {\bibinfo  {journal} {Nature}\ }\textbf {\bibinfo {volume} {573}},\
  \bibinfo {pages} {91} (\bibinfo {year} {2019})}\BibitemShut {NoStop}%
\bibitem [{\citenamefont {Choi}\ \emph {et~al.}(2019)\citenamefont {Choi},
  \citenamefont {Kemmer}, \citenamefont {Peng}, \citenamefont {Thomson},
  \citenamefont {Arora}, \citenamefont {Polski}, \citenamefont {Zhang},
  \citenamefont {Ren}, \citenamefont {Alicea}, \citenamefont {Refael},
  \citenamefont {von Oppen}, \citenamefont {Watanabe}, \citenamefont
  {Taniguchi},\ and\ \citenamefont
  {Nadj-Perge}}]{Electronic-correlations-2019}%
  \BibitemOpen
  \bibfield  {author} {\bibinfo {author} {\bibfnamefont {Y.}~\bibnamefont
  {Choi}}, \bibinfo {author} {\bibfnamefont {J.}~\bibnamefont {Kemmer}},
  \bibinfo {author} {\bibfnamefont {Y.}~\bibnamefont {Peng}}, \bibinfo {author}
  {\bibfnamefont {A.}~\bibnamefont {Thomson}}, \bibinfo {author} {\bibfnamefont
  {H.}~\bibnamefont {Arora}}, \bibinfo {author} {\bibfnamefont
  {R.}~\bibnamefont {Polski}}, \bibinfo {author} {\bibfnamefont
  {Y.}~\bibnamefont {Zhang}}, \bibinfo {author} {\bibfnamefont
  {H.}~\bibnamefont {Ren}}, \bibinfo {author} {\bibfnamefont {J.}~\bibnamefont
  {Alicea}}, \bibinfo {author} {\bibfnamefont {G.}~\bibnamefont {Refael}},
  \bibinfo {author} {\bibfnamefont {F.}~\bibnamefont {von Oppen}}, \bibinfo
  {author} {\bibfnamefont {K.}~\bibnamefont {Watanabe}}, \bibinfo {author}
  {\bibfnamefont {T.}~\bibnamefont {Taniguchi}}, \ and\ \bibinfo {author}
  {\bibfnamefont {S.}~\bibnamefont {Nadj-Perge}},\ }\bibfield  {title}
  {\enquote {\bibinfo {title} {{E}lectronic correlations in twisted bilayer
  graphene near the magic angle},}\ }\href {\doibase 10.1038/s41567-019-0606-5}
  {\bibfield  {journal} {\bibinfo  {journal} {Nat. Phys.}\ }\textbf {\bibinfo
  {volume} {15}},\ \bibinfo {pages} {1174} (\bibinfo {year}
  {2019})}\BibitemShut {NoStop}%
\bibitem [{\citenamefont {Yankowitz}\ \emph {et~al.}(2019)\citenamefont
  {Yankowitz}, \citenamefont {Chen}, \citenamefont {Polshyn}, \citenamefont
  {Zhang}, \citenamefont {Watanabe}, \citenamefont {Taniguchi}, \citenamefont
  {Graf}, \citenamefont {Young},\ and\ \citenamefont
  {Dean}}]{TuningYankowitz1059}%
  \BibitemOpen
  \bibfield  {author} {\bibinfo {author} {\bibfnamefont {M.}~\bibnamefont
  {Yankowitz}}, \bibinfo {author} {\bibfnamefont {S.}~\bibnamefont {Chen}},
  \bibinfo {author} {\bibfnamefont {H.}~\bibnamefont {Polshyn}}, \bibinfo
  {author} {\bibfnamefont {Y.}~\bibnamefont {Zhang}}, \bibinfo {author}
  {\bibfnamefont {K.}~\bibnamefont {Watanabe}}, \bibinfo {author}
  {\bibfnamefont {T.}~\bibnamefont {Taniguchi}}, \bibinfo {author}
  {\bibfnamefont {D.}~\bibnamefont {Graf}}, \bibinfo {author} {\bibfnamefont
  {A.~F.}\ \bibnamefont {Young}}, \ and\ \bibinfo {author} {\bibfnamefont
  {C.~R.}\ \bibnamefont {Dean}},\ }\bibfield  {title} {\enquote {\bibinfo
  {title} {Tuning superconductivity in twisted bilayer graphene},}\ }\href
  {\doibase 10.1126/science.aav1910} {\bibfield  {journal} {\bibinfo  {journal}
  {Science}\ }\textbf {\bibinfo {volume} {363}},\ \bibinfo {pages} {1059}
  (\bibinfo {year} {2019})}\BibitemShut {NoStop}%
\bibitem [{\citenamefont {Uri}\ \emph {et~al.}(2020)\citenamefont {Uri},
  \citenamefont {Grover}, \citenamefont {Cao}, \citenamefont {Crosse},
  \citenamefont {Bagani}, \citenamefont {Rodan-Legrain}, \citenamefont
  {Myasoedov}, \citenamefont {Watanabe}, \citenamefont {Taniguchi},
  \citenamefont {Moon}, \citenamefont {Koshino}, \citenamefont
  {Jarillo-Herrero},\ and\ \citenamefont {Zeldov}}]{uri2020mapping}%
  \BibitemOpen
  \bibfield  {author} {\bibinfo {author} {\bibfnamefont {A.}~\bibnamefont
  {Uri}}, \bibinfo {author} {\bibfnamefont {S.}~\bibnamefont {Grover}},
  \bibinfo {author} {\bibfnamefont {Y.}~\bibnamefont {Cao}}, \bibinfo {author}
  {\bibfnamefont {J.A.}\ \bibnamefont {Crosse}}, \bibinfo {author}
  {\bibfnamefont {K.}~\bibnamefont {Bagani}}, \bibinfo {author} {\bibfnamefont
  {D.}~\bibnamefont {Rodan-Legrain}}, \bibinfo {author} {\bibfnamefont
  {Y.}~\bibnamefont {Myasoedov}}, \bibinfo {author} {\bibfnamefont
  {K.}~\bibnamefont {Watanabe}}, \bibinfo {author} {\bibfnamefont
  {T.}~\bibnamefont {Taniguchi}}, \bibinfo {author} {\bibfnamefont
  {P.}~\bibnamefont {Moon}}, \bibinfo {author} {\bibfnamefont {M.}~\bibnamefont
  {Koshino}}, \bibinfo {author} {\bibfnamefont {P.}~\bibnamefont
  {Jarillo-Herrero}}, \ and\ \bibinfo {author} {\bibfnamefont {E.}~\bibnamefont
  {Zeldov}},\ }\bibfield  {title} {\enquote {\bibinfo {title} {Mapping the
  twist-angle disorder and {L}andau levels in magic-angle graphene},}\ }\href
  {https://www.nature.com/articles/s41586-020-2255-3} {\bibfield  {journal}
  {\bibinfo  {journal} {Nature}\ }\textbf {\bibinfo {volume} {581}},\ \bibinfo
  {pages} {47} (\bibinfo {year} {2020})}\BibitemShut {NoStop}%
\bibitem [{\citenamefont {Nuckolls}\ \emph {et~al.}(2020)\citenamefont
  {Nuckolls}, \citenamefont {Oh}, \citenamefont {Wong}, \citenamefont {Lian},
  \citenamefont {Watanabe}, \citenamefont {Taniguchi}, \citenamefont
  {Bernevig},\ and\ \citenamefont {Yazdani}}]{Strongly-correlated-2020}%
  \BibitemOpen
  \bibfield  {author} {\bibinfo {author} {\bibfnamefont {K.~P.}\ \bibnamefont
  {Nuckolls}}, \bibinfo {author} {\bibfnamefont {M.}~\bibnamefont {Oh}},
  \bibinfo {author} {\bibfnamefont {D.}~\bibnamefont {Wong}}, \bibinfo {author}
  {\bibfnamefont {B.}~\bibnamefont {Lian}}, \bibinfo {author} {\bibfnamefont
  {K.}~\bibnamefont {Watanabe}}, \bibinfo {author} {\bibfnamefont
  {T.}~\bibnamefont {Taniguchi}}, \bibinfo {author} {\bibfnamefont {B.~A.}\
  \bibnamefont {Bernevig}}, \ and\ \bibinfo {author} {\bibfnamefont
  {A.}~\bibnamefont {Yazdani}},\ }\bibfield  {title} {\enquote {\bibinfo
  {title} {Strongly correlated {C}hern insulators in magic-angle twisted
  bilayer graphene},}\ }\href {\doibase 10.1038/s41586-020-3028-8} {\bibfield
  {journal} {\bibinfo  {journal} {Nature}\ }\textbf {\bibinfo {volume} {588}},\
  \bibinfo {pages} {610} (\bibinfo {year} {2020})}\BibitemShut {NoStop}%
\bibitem [{\citenamefont {Wong}\ \emph {et~al.}(2020)\citenamefont {Wong},
  \citenamefont {Nuckolls}, \citenamefont {Oh}, \citenamefont {Lian},
  \citenamefont {Xie}, \citenamefont {Jeon}, \citenamefont {Watanabe},
  \citenamefont {Taniguchi}, \citenamefont {Bernevig},\ and\ \citenamefont
  {Yazdani}}]{Cascade-of-e-2020}%
  \BibitemOpen
  \bibfield  {author} {\bibinfo {author} {\bibfnamefont {D.}~\bibnamefont
  {Wong}}, \bibinfo {author} {\bibfnamefont {K.~P.}\ \bibnamefont {Nuckolls}},
  \bibinfo {author} {\bibfnamefont {M.}~\bibnamefont {Oh}}, \bibinfo {author}
  {\bibfnamefont {B.}~\bibnamefont {Lian}}, \bibinfo {author} {\bibfnamefont
  {Y.}~\bibnamefont {Xie}}, \bibinfo {author} {\bibfnamefont {S.}~\bibnamefont
  {Jeon}}, \bibinfo {author} {\bibfnamefont {K.}~\bibnamefont {Watanabe}},
  \bibinfo {author} {\bibfnamefont {T.}~\bibnamefont {Taniguchi}}, \bibinfo
  {author} {\bibfnamefont {B.~A.}\ \bibnamefont {Bernevig}}, \ and\ \bibinfo
  {author} {\bibfnamefont {A.}~\bibnamefont {Yazdani}},\ }\bibfield  {title}
  {\enquote {\bibinfo {title} {{C}ascade of electronic transitions in
  magic-angle twisted bilayer graphene},}\ }\href {\doibase
  10.1038/s41586-020-2339-0} {\bibfield  {journal} {\bibinfo  {journal}
  {Nature}\ }\textbf {\bibinfo {volume} {582}},\ \bibinfo {pages} {198}
  (\bibinfo {year} {2020})}\BibitemShut {NoStop}%
\bibitem [{\citenamefont {Zondiner}\ \emph {et~al.}(2020)\citenamefont
  {Zondiner}, \citenamefont {Rozen}, \citenamefont {Rodan-Legrain},
  \citenamefont {Cao}, \citenamefont {Queiroz}, \citenamefont {Taniguchi},
  \citenamefont {Watanabe}, \citenamefont {Oreg}, \citenamefont {von Oppen},
  \citenamefont {Stern}, \citenamefont {Berg}, \citenamefont
  {Jarillo-Herrero},\ and\ \citenamefont {Ilani}}]{Cascade-of-p-2020}%
  \BibitemOpen
  \bibfield  {author} {\bibinfo {author} {\bibfnamefont {U.}~\bibnamefont
  {Zondiner}}, \bibinfo {author} {\bibfnamefont {A.}~\bibnamefont {Rozen}},
  \bibinfo {author} {\bibfnamefont {D.}~\bibnamefont {Rodan-Legrain}}, \bibinfo
  {author} {\bibfnamefont {Y.}~\bibnamefont {Cao}}, \bibinfo {author}
  {\bibfnamefont {R.}~\bibnamefont {Queiroz}}, \bibinfo {author} {\bibfnamefont
  {T.}~\bibnamefont {Taniguchi}}, \bibinfo {author} {\bibfnamefont
  {K.}~\bibnamefont {Watanabe}}, \bibinfo {author} {\bibfnamefont
  {Y.}~\bibnamefont {Oreg}}, \bibinfo {author} {\bibfnamefont {F.}~\bibnamefont
  {von Oppen}}, \bibinfo {author} {\bibfnamefont {A.}~\bibnamefont {Stern}},
  \bibinfo {author} {\bibfnamefont {E.}~\bibnamefont {Berg}}, \bibinfo {author}
  {\bibfnamefont {P.}~\bibnamefont {Jarillo-Herrero}}, \ and\ \bibinfo {author}
  {\bibfnamefont {S.}~\bibnamefont {Ilani}},\ }\bibfield  {title} {\enquote
  {\bibinfo {title} {{C}ascade of phase transitions and {D}irac revivals in
  magic-angle graphene},}\ }\href {\doibase 10.1038/s41586-020-2373-y}
  {\bibfield  {journal} {\bibinfo  {journal} {Nature}\ }\textbf {\bibinfo
  {volume} {582}},\ \bibinfo {pages} {203} (\bibinfo {year}
  {2020})}\BibitemShut {NoStop}%
\bibitem [{\citenamefont {Saito}\ \emph {et~al.}(2020)\citenamefont {Saito},
  \citenamefont {Ge}, \citenamefont {Watanabe}, \citenamefont {Taniguchi},\
  and\ \citenamefont {Young}}]{Independent-superconductors-2020}%
  \BibitemOpen
  \bibfield  {author} {\bibinfo {author} {\bibfnamefont {Y.}~\bibnamefont
  {Saito}}, \bibinfo {author} {\bibfnamefont {J.}~\bibnamefont {Ge}}, \bibinfo
  {author} {\bibfnamefont {K.}~\bibnamefont {Watanabe}}, \bibinfo {author}
  {\bibfnamefont {T.}~\bibnamefont {Taniguchi}}, \ and\ \bibinfo {author}
  {\bibfnamefont {A.~F.}\ \bibnamefont {Young}},\ }\bibfield  {title} {\enquote
  {\bibinfo {title} {{I}ndependent superconductors and correlated insulators in
  twisted bilayer graphene},}\ }\href {\doibase 10.1038/s41567-020-0928-3}
  {\bibfield  {journal} {\bibinfo  {journal} {Nat. Phys.}\ }\textbf {\bibinfo
  {volume} {16}},\ \bibinfo {pages} {926} (\bibinfo {year} {2020})}\BibitemShut
  {NoStop}%
\bibitem [{\citenamefont {Stepanov}\ \emph {et~al.}(2020)\citenamefont
  {Stepanov}, \citenamefont {Das}, \citenamefont {Lu}, \citenamefont
  {Fahimniya}, \citenamefont {Watanabe}, \citenamefont {Taniguchi},
  \citenamefont {Koppens}, \citenamefont {Lischner}, \citenamefont {Levitov},\
  and\ \citenamefont {Efetov}}]{Untying-the-2020}%
  \BibitemOpen
  \bibfield  {author} {\bibinfo {author} {\bibfnamefont {P.}~\bibnamefont
  {Stepanov}}, \bibinfo {author} {\bibfnamefont {I.}~\bibnamefont {Das}},
  \bibinfo {author} {\bibfnamefont {X.}~\bibnamefont {Lu}}, \bibinfo {author}
  {\bibfnamefont {A.}~\bibnamefont {Fahimniya}}, \bibinfo {author}
  {\bibfnamefont {K.}~\bibnamefont {Watanabe}}, \bibinfo {author}
  {\bibfnamefont {T.}~\bibnamefont {Taniguchi}}, \bibinfo {author}
  {\bibfnamefont {F.~H.~L.}\ \bibnamefont {Koppens}}, \bibinfo {author}
  {\bibfnamefont {J.}~\bibnamefont {Lischner}}, \bibinfo {author}
  {\bibfnamefont {L.}~\bibnamefont {Levitov}}, \ and\ \bibinfo {author}
  {\bibfnamefont {D.~K.}\ \bibnamefont {Efetov}},\ }\bibfield  {title}
  {\enquote {\bibinfo {title} {{U}ntying the insulating and superconducting
  orders in magic-angle graphene},}\ }\href {\doibase
  10.1038/s41586-020-2459-6} {\bibfield  {journal} {\bibinfo  {journal}
  {Nature}\ }\textbf {\bibinfo {volume} {583}},\ \bibinfo {pages} {375}
  (\bibinfo {year} {2020})}\BibitemShut {NoStop}%
\bibitem [{\citenamefont {Luo}\ \emph {et~al.}(2020)\citenamefont {Luo},
  \citenamefont {Engelke}, \citenamefont {Mattheakis}, \citenamefont
  {Tamagnone}, \citenamefont {Carr}, \citenamefont {Watanabe}, \citenamefont
  {Taniguchi}, \citenamefont {Kaxiras}, \citenamefont {Kim},\ and\
  \citenamefont {Wilson}}]{In-situ-2020}%
  \BibitemOpen
  \bibfield  {author} {\bibinfo {author} {\bibfnamefont {Y.}~\bibnamefont
  {Luo}}, \bibinfo {author} {\bibfnamefont {R.}~\bibnamefont {Engelke}},
  \bibinfo {author} {\bibfnamefont {M.}~\bibnamefont {Mattheakis}}, \bibinfo
  {author} {\bibfnamefont {M.}~\bibnamefont {Tamagnone}}, \bibinfo {author}
  {\bibfnamefont {S.}~\bibnamefont {Carr}}, \bibinfo {author} {\bibfnamefont
  {K.}~\bibnamefont {Watanabe}}, \bibinfo {author} {\bibfnamefont
  {T.}~\bibnamefont {Taniguchi}}, \bibinfo {author} {\bibfnamefont
  {E.}~\bibnamefont {Kaxiras}}, \bibinfo {author} {\bibfnamefont
  {P.}~\bibnamefont {Kim}}, \ and\ \bibinfo {author} {\bibfnamefont {W.~L.}\
  \bibnamefont {Wilson}},\ }\bibfield  {title} {\enquote {\bibinfo {title}
  {{I}n situ nanoscale imaging of moir\'e superlattices in twisted van der
  {W}aals heterostructures},}\ }\href {\doibase 10.1038/s41467-020-18109-0}
  {\bibfield  {journal} {\bibinfo  {journal} {Nat. Commun.}\ }\textbf {\bibinfo
  {volume} {11}},\ \bibinfo {pages} {4209} (\bibinfo {year}
  {2020})}\BibitemShut {NoStop}%
\bibitem [{\citenamefont {Rozen}\ \emph {et~al.}(2021)\citenamefont {Rozen},
  \citenamefont {Park}, \citenamefont {Zondiner}, \citenamefont {Cao},
  \citenamefont {Rodan-Legrain}, \citenamefont {Taniguchi}, \citenamefont
  {Watanabe}, \citenamefont {Oreg}, \citenamefont {Stern}, \citenamefont
  {Berg}, \citenamefont {Jarillo-Herrero},\ and\ \citenamefont
  {Ilani}}]{Entropic-evidence-2021}%
  \BibitemOpen
  \bibfield  {author} {\bibinfo {author} {\bibfnamefont {A.}~\bibnamefont
  {Rozen}}, \bibinfo {author} {\bibfnamefont {J.~M.}\ \bibnamefont {Park}},
  \bibinfo {author} {\bibfnamefont {U.}~\bibnamefont {Zondiner}}, \bibinfo
  {author} {\bibfnamefont {Y.}~\bibnamefont {Cao}}, \bibinfo {author}
  {\bibfnamefont {D.}~\bibnamefont {Rodan-Legrain}}, \bibinfo {author}
  {\bibfnamefont {T.}~\bibnamefont {Taniguchi}}, \bibinfo {author}
  {\bibfnamefont {K.}~\bibnamefont {Watanabe}}, \bibinfo {author}
  {\bibfnamefont {Y.}~\bibnamefont {Oreg}}, \bibinfo {author} {\bibfnamefont
  {A.}~\bibnamefont {Stern}}, \bibinfo {author} {\bibfnamefont
  {E.}~\bibnamefont {Berg}}, \bibinfo {author} {\bibfnamefont {P.}~\bibnamefont
  {Jarillo-Herrero}}, \ and\ \bibinfo {author} {\bibfnamefont {S.}~\bibnamefont
  {Ilani}},\ }\bibfield  {title} {\enquote {\bibinfo {title} {{E}ntropic
  evidence for a {P}omeranchuk effect in magic-angle graphene},}\ }\href
  {\doibase 10.1038/s41586-021-03319-3} {\bibfield  {journal} {\bibinfo
  {journal} {Nature}\ }\textbf {\bibinfo {volume} {592}},\ \bibinfo {pages}
  {214} (\bibinfo {year} {2021})}\BibitemShut {NoStop}%
\bibitem [{\citenamefont {Hesp}\ \emph {et~al.}(2021)\citenamefont {Hesp},
  \citenamefont {Torre}, \citenamefont {Rodan-Legrain}, \citenamefont
  {Novelli}, \citenamefont {Cao}, \citenamefont {Carr}, \citenamefont {Fang},
  \citenamefont {Stepanov}, \citenamefont {Barcons-Ruiz}, \citenamefont
  {Herzig~Sheinfux}, \citenamefont {Watanabe}, \citenamefont {Taniguchi},
  \citenamefont {Efetov}, \citenamefont {Kaxiras}, \citenamefont
  {Jarillo-Herrero}, \citenamefont {Polini},\ and\ \citenamefont
  {Koppens}}]{Observation-of-2021}%
  \BibitemOpen
  \bibfield  {author} {\bibinfo {author} {\bibfnamefont {N.~C.~H.}\
  \bibnamefont {Hesp}}, \bibinfo {author} {\bibfnamefont {I.}~\bibnamefont
  {Torre}}, \bibinfo {author} {\bibfnamefont {D.}~\bibnamefont
  {Rodan-Legrain}}, \bibinfo {author} {\bibfnamefont {P.}~\bibnamefont
  {Novelli}}, \bibinfo {author} {\bibfnamefont {Y.}~\bibnamefont {Cao}},
  \bibinfo {author} {\bibfnamefont {S.}~\bibnamefont {Carr}}, \bibinfo {author}
  {\bibfnamefont {S.}~\bibnamefont {Fang}}, \bibinfo {author} {\bibfnamefont
  {P.}~\bibnamefont {Stepanov}}, \bibinfo {author} {\bibfnamefont
  {D.}~\bibnamefont {Barcons-Ruiz}}, \bibinfo {author} {\bibfnamefont
  {H.}~\bibnamefont {Herzig~Sheinfux}}, \bibinfo {author} {\bibfnamefont
  {K.}~\bibnamefont {Watanabe}}, \bibinfo {author} {\bibfnamefont
  {T.}~\bibnamefont {Taniguchi}}, \bibinfo {author} {\bibfnamefont {D.~K.}\
  \bibnamefont {Efetov}}, \bibinfo {author} {\bibfnamefont {E.}~\bibnamefont
  {Kaxiras}}, \bibinfo {author} {\bibfnamefont {P.}~\bibnamefont
  {Jarillo-Herrero}}, \bibinfo {author} {\bibfnamefont {M.}~\bibnamefont
  {Polini}}, \ and\ \bibinfo {author} {\bibfnamefont {F.~H.~L.}\ \bibnamefont
  {Koppens}},\ }\bibfield  {title} {\enquote {\bibinfo {title} {{O}bservation
  of interband collective excitations in twisted bilayer graphene},}\ }\href
  {\doibase 10.1038/s41567-021-01327-8} {\bibfield  {journal} {\bibinfo
  {journal} {Nat. Phys.}\ }\textbf {\bibinfo {volume} {17}},\ \bibinfo {pages}
  {1162} (\bibinfo {year} {2021})}\BibitemShut {NoStop}%
\bibitem [{\citenamefont {Ma}\ \emph {et~al.}(2022)\citenamefont {Ma},
  \citenamefont {Yuan}, \citenamefont {Cheung}, \citenamefont {Watanabe},
  \citenamefont {Taniguchi}, \citenamefont {Zhang},\ and\ \citenamefont
  {Xia}}]{Intelligent-infrared-2022}%
  \BibitemOpen
  \bibfield  {author} {\bibinfo {author} {\bibfnamefont {C.}~\bibnamefont
  {Ma}}, \bibinfo {author} {\bibfnamefont {S.}~\bibnamefont {Yuan}}, \bibinfo
  {author} {\bibfnamefont {P.}~\bibnamefont {Cheung}}, \bibinfo {author}
  {\bibfnamefont {K.}~\bibnamefont {Watanabe}}, \bibinfo {author}
  {\bibfnamefont {T.}~\bibnamefont {Taniguchi}}, \bibinfo {author}
  {\bibfnamefont {F.}~\bibnamefont {Zhang}}, \ and\ \bibinfo {author}
  {\bibfnamefont {F.}~\bibnamefont {Xia}},\ }\bibfield  {title} {\enquote
  {\bibinfo {title} {{I}ntelligent infrared sensing enabled by tunable moir\'e
  quantum geometry},}\ }\href {\doibase 10.1038/s41586-022-04548-w} {\bibfield
  {journal} {\bibinfo  {journal} {Nature}\ }\textbf {\bibinfo {volume} {604}},\
  \bibinfo {pages} {266} (\bibinfo {year} {2022})}\BibitemShut {NoStop}%
\bibitem [{\citenamefont {Carr}\ \emph {et~al.}(2020)\citenamefont {Carr},
  \citenamefont {Fang},\ and\ \citenamefont
  {Kaxiras}}]{Electronic-structure-methods-2020}%
  \BibitemOpen
  \bibfield  {author} {\bibinfo {author} {\bibfnamefont {S.}~\bibnamefont
  {Carr}}, \bibinfo {author} {\bibfnamefont {S.}~\bibnamefont {Fang}}, \ and\
  \bibinfo {author} {\bibfnamefont {E.}~\bibnamefont {Kaxiras}},\ }\bibfield
  {title} {\enquote {\bibinfo {title} {{E}lectronic-structure methods for
  twisted moir\'{e} layers},}\ }\href {\doibase 10.1038/s41578-020-0214-0}
  {\bibfield  {journal} {\bibinfo  {journal} {Nat. Rev. Mater.}\ }\textbf
  {\bibinfo {volume} {5}},\ \bibinfo {pages} {748} (\bibinfo {year}
  {2020})}\BibitemShut {NoStop}%
\bibitem [{\citenamefont {Andrei}\ and\ \citenamefont
  {MacDonald}(2020)}]{Graphene-bilayers-2020}%
  \BibitemOpen
  \bibfield  {author} {\bibinfo {author} {\bibfnamefont {E.~Y.}\ \bibnamefont
  {Andrei}}\ and\ \bibinfo {author} {\bibfnamefont {A.~H.}\ \bibnamefont
  {MacDonald}},\ }\bibfield  {title} {\enquote {\bibinfo {title} {Graphene
  bilayers with a twist},}\ }\href {\doibase 10.1038/s41563-020-00840-0}
  {\bibfield  {journal} {\bibinfo  {journal} {Nat. Mater.}\ }\textbf {\bibinfo
  {volume} {19}},\ \bibinfo {pages} {1265} (\bibinfo {year}
  {2020})}\BibitemShut {NoStop}%
\bibitem [{\citenamefont {Andrei}\ \emph {et~al.}(2021)\citenamefont {Andrei},
  \citenamefont {Efetov}, \citenamefont {Jarillo-Herrero}, \citenamefont
  {MacDonald}, \citenamefont {Mak}, \citenamefont {Senthil}, \citenamefont
  {Tutuc}, \citenamefont {Yazdani},\ and\ \citenamefont
  {Young}}]{The-marvels-2021}%
  \BibitemOpen
  \bibfield  {author} {\bibinfo {author} {\bibfnamefont {E.~Y.}\ \bibnamefont
  {Andrei}}, \bibinfo {author} {\bibfnamefont {D.~K.}\ \bibnamefont {Efetov}},
  \bibinfo {author} {\bibfnamefont {P.}~\bibnamefont {Jarillo-Herrero}},
  \bibinfo {author} {\bibfnamefont {A.~H.}\ \bibnamefont {MacDonald}}, \bibinfo
  {author} {\bibfnamefont {K.~F.}\ \bibnamefont {Mak}}, \bibinfo {author}
  {\bibfnamefont {T.}~\bibnamefont {Senthil}}, \bibinfo {author} {\bibfnamefont
  {E.}~\bibnamefont {Tutuc}}, \bibinfo {author} {\bibfnamefont
  {A.}~\bibnamefont {Yazdani}}, \ and\ \bibinfo {author} {\bibfnamefont
  {A.~F.}\ \bibnamefont {Young}},\ }\bibfield  {title} {\enquote {\bibinfo
  {title} {The marvels of moir\'{e} materials},}\ }\href {\doibase
  10.1038/s41578-021-00284-1} {\bibfield  {journal} {\bibinfo  {journal} {Nat.
  Rev. Mater.}\ }\textbf {\bibinfo {volume} {6}},\ \bibinfo {pages} {201}
  (\bibinfo {year} {2021})}\BibitemShut {NoStop}%
\bibitem [{\citenamefont {Liu}\ and\ \citenamefont
  {Dai}(2021{\natexlab{a}})}]{Orbital-magnetic-2021}%
  \BibitemOpen
  \bibfield  {author} {\bibinfo {author} {\bibfnamefont {J.}~\bibnamefont
  {Liu}}\ and\ \bibinfo {author} {\bibfnamefont {X.}~\bibnamefont {Dai}},\
  }\bibfield  {title} {\enquote {\bibinfo {title} {{O}rbital magnetic states in
  moir\'{e} graphene systems},}\ }\href {\doibase 10.1038/s42254-021-00297-3}
  {\bibfield  {journal} {\bibinfo  {journal} {Nat. Rev. Phys.}\ }\textbf
  {\bibinfo {volume} {3}},\ \bibinfo {pages} {367} (\bibinfo {year}
  {2021}{\natexlab{a}})}\BibitemShut {NoStop}%
\bibitem [{\citenamefont {Po}\ \emph {et~al.}(2018)\citenamefont {Po},
  \citenamefont {Zou}, \citenamefont {Vishwanath},\ and\ \citenamefont
  {Senthil}}]{Origin-of-mott2018}%
  \BibitemOpen
  \bibfield  {author} {\bibinfo {author} {\bibfnamefont {H.~C.}\ \bibnamefont
  {Po}}, \bibinfo {author} {\bibfnamefont {L.}~\bibnamefont {Zou}}, \bibinfo
  {author} {\bibfnamefont {A.}~\bibnamefont {Vishwanath}}, \ and\ \bibinfo
  {author} {\bibfnamefont {T.}~\bibnamefont {Senthil}},\ }\bibfield  {title}
  {\enquote {\bibinfo {title} {{O}rigin of {M}ott {I}nsulating {B}ehavior and
  {S}uperconductivity in {T}wisted {B}ilayer {G}raphene},}\ }\href {\doibase
  10.1103/PhysRevX.8.031089} {\bibfield  {journal} {\bibinfo  {journal} {Phys.
  Rev. X}\ }\textbf {\bibinfo {volume} {8}},\ \bibinfo {pages} {031089}
  (\bibinfo {year} {2018})}\BibitemShut {NoStop}%
\bibitem [{\citenamefont {Koshino}\ \emph {et~al.}(2018)\citenamefont
  {Koshino}, \citenamefont {Yuan}, \citenamefont {Koretsune}, \citenamefont
  {Ochi}, \citenamefont {Kuroki},\ and\ \citenamefont
  {Fu}}]{Maximally-Localized-2018}%
  \BibitemOpen
  \bibfield  {author} {\bibinfo {author} {\bibfnamefont {M.}~\bibnamefont
  {Koshino}}, \bibinfo {author} {\bibfnamefont {N.~F.~Q.}\ \bibnamefont
  {Yuan}}, \bibinfo {author} {\bibfnamefont {T.}~\bibnamefont {Koretsune}},
  \bibinfo {author} {\bibfnamefont {M.}~\bibnamefont {Ochi}}, \bibinfo {author}
  {\bibfnamefont {K.}~\bibnamefont {Kuroki}}, \ and\ \bibinfo {author}
  {\bibfnamefont {L.}~\bibnamefont {Fu}},\ }\bibfield  {title} {\enquote
  {\bibinfo {title} {{M}aximally {L}ocalized {W}annier {O}rbitals and the
  {E}xtended {H}ubbard {M}odel for {T}wisted {B}ilayer {G}raphene},}\ }\href
  {\doibase 10.1103/PhysRevX.8.031087} {\bibfield  {journal} {\bibinfo
  {journal} {Phys. Rev. X}\ }\textbf {\bibinfo {volume} {8}},\ \bibinfo {pages}
  {031087} (\bibinfo {year} {2018})}\BibitemShut {NoStop}%
\bibitem [{\citenamefont {Kang}\ and\ \citenamefont
  {Vafek}(2018)}]{Symmetry-Maximally-2018}%
  \BibitemOpen
  \bibfield  {author} {\bibinfo {author} {\bibfnamefont {J.}~\bibnamefont
  {Kang}}\ and\ \bibinfo {author} {\bibfnamefont {O.}~\bibnamefont {Vafek}},\
  }\bibfield  {title} {\enquote {\bibinfo {title} {{S}ymmetry, {M}aximally
  {L}ocalized {W}annier {S}tates, and a {L}ow-{E}nergy {M}odel for {T}wisted
  {B}ilayer {G}raphene {N}arrow {B}ands},}\ }\href {\doibase
  10.1103/PhysRevX.8.031088} {\bibfield  {journal} {\bibinfo  {journal} {Phys.
  Rev. X}\ }\textbf {\bibinfo {volume} {8}},\ \bibinfo {pages} {031088}
  (\bibinfo {year} {2018})}\BibitemShut {NoStop}%
\bibitem [{\citenamefont {Isobe}\ \emph {et~al.}(2018)\citenamefont {Isobe},
  \citenamefont {Yuan},\ and\ \citenamefont
  {Fu}}]{Unconventional-Supercond-2018}%
  \BibitemOpen
  \bibfield  {author} {\bibinfo {author} {\bibfnamefont {H.}~\bibnamefont
  {Isobe}}, \bibinfo {author} {\bibfnamefont {N.~F.~Q.}\ \bibnamefont {Yuan}},
  \ and\ \bibinfo {author} {\bibfnamefont {L.}~\bibnamefont {Fu}},\ }\bibfield
  {title} {\enquote {\bibinfo {title} {{U}nconventional {S}uperconductivity and
  {D}ensity {W}aves in {T}wisted {B}ilayer {G}raphene},}\ }\href {\doibase
  10.1103/PhysRevX.8.041041} {\bibfield  {journal} {\bibinfo  {journal} {Phys.
  Rev. X}\ }\textbf {\bibinfo {volume} {8}},\ \bibinfo {pages} {041041}
  (\bibinfo {year} {2018})}\BibitemShut {NoStop}%
\bibitem [{\citenamefont {Liu}\ \emph {et~al.}(2018)\citenamefont {Liu},
  \citenamefont {Zhang}, \citenamefont {Chen},\ and\ \citenamefont
  {Yang}}]{Chiral-Spin-2018}%
  \BibitemOpen
  \bibfield  {author} {\bibinfo {author} {\bibfnamefont {C.-C.}\ \bibnamefont
  {Liu}}, \bibinfo {author} {\bibfnamefont {L.-D.}\ \bibnamefont {Zhang}},
  \bibinfo {author} {\bibfnamefont {W.-Q.}\ \bibnamefont {Chen}}, \ and\
  \bibinfo {author} {\bibfnamefont {F.}~\bibnamefont {Yang}},\ }\bibfield
  {title} {\enquote {\bibinfo {title} {{C}hiral {S}pin {D}ensity {W}ave and
  $d+id$ {S}uperconductivity in the {M}agic-{A}ngle-{T}wisted {B}ilayer
  {G}raphene},}\ }\href {\doibase 10.1103/PhysRevLett.121.217001} {\bibfield
  {journal} {\bibinfo  {journal} {Phys. Rev. Lett.}\ }\textbf {\bibinfo
  {volume} {121}},\ \bibinfo {pages} {217001} (\bibinfo {year}
  {2018})}\BibitemShut {NoStop}%
\bibitem [{\citenamefont {Kang}\ and\ \citenamefont
  {Vafek}(2019)}]{Strong-Coupl-2019}%
  \BibitemOpen
  \bibfield  {author} {\bibinfo {author} {\bibfnamefont {J.}~\bibnamefont
  {Kang}}\ and\ \bibinfo {author} {\bibfnamefont {O.}~\bibnamefont {Vafek}},\
  }\bibfield  {title} {\enquote {\bibinfo {title} {{S}trong {C}oupling {P}hases
  of {P}artially {F}illed {T}wisted {B}ilayer {G}raphene {N}arrow {B}ands},}\
  }\href {\doibase 10.1103/PhysRevLett.122.246401} {\bibfield  {journal}
  {\bibinfo  {journal} {Phys. Rev. Lett.}\ }\textbf {\bibinfo {volume} {122}},\
  \bibinfo {pages} {246401} (\bibinfo {year} {2019})}\BibitemShut {NoStop}%
\bibitem [{\citenamefont {Zhang}\ and\ \citenamefont
  {Senthil}(2019)}]{Bridging-Hubbard-2019}%
  \BibitemOpen
  \bibfield  {author} {\bibinfo {author} {\bibfnamefont {Y.-H.}\ \bibnamefont
  {Zhang}}\ and\ \bibinfo {author} {\bibfnamefont {T.}~\bibnamefont
  {Senthil}},\ }\bibfield  {title} {\enquote {\bibinfo {title} {{B}ridging
  {H}ubbard model physics and quantum {H}all physics in trilayer
  $\text{graphene}/h\ensuremath{-}\mathrm{{B}{N}}$ moir\'e superlattice},}\
  }\href {\doibase 10.1103/PhysRevB.99.205150} {\bibfield  {journal} {\bibinfo
  {journal} {Phys. Rev. B}\ }\textbf {\bibinfo {volume} {99}},\ \bibinfo
  {pages} {205150} (\bibinfo {year} {2019})}\BibitemShut {NoStop}%
\bibitem [{\citenamefont {Chittari}\ \emph {et~al.}(2019)\citenamefont
  {Chittari}, \citenamefont {Chen}, \citenamefont {Zhang}, \citenamefont
  {Wang},\ and\ \citenamefont {Jung}}]{Gate-TunableTopological-2019}%
  \BibitemOpen
  \bibfield  {author} {\bibinfo {author} {\bibfnamefont {B.~L.}\ \bibnamefont
  {Chittari}}, \bibinfo {author} {\bibfnamefont {G.}~\bibnamefont {Chen}},
  \bibinfo {author} {\bibfnamefont {Y.}~\bibnamefont {Zhang}}, \bibinfo
  {author} {\bibfnamefont {F.}~\bibnamefont {Wang}}, \ and\ \bibinfo {author}
  {\bibfnamefont {J.}~\bibnamefont {Jung}},\ }\bibfield  {title} {\enquote
  {\bibinfo {title} {{G}ate-{T}unable {T}opological {F}lat {B}ands in
  {T}rilayer {G}raphene {B}oron-{N}itride {M}oir\'e {S}uperlattices},}\ }\href
  {\doibase 10.1103/PhysRevLett.122.016401} {\bibfield  {journal} {\bibinfo
  {journal} {Phys. Rev. Lett.}\ }\textbf {\bibinfo {volume} {122}},\ \bibinfo
  {pages} {016401} (\bibinfo {year} {2019})}\BibitemShut {NoStop}%
\bibitem [{\citenamefont {Zhang}\ \emph {et~al.}(2019)\citenamefont {Zhang},
  \citenamefont {Mao},\ and\ \citenamefont {Senthil}}]{Twisted209zhang}%
  \BibitemOpen
  \bibfield  {author} {\bibinfo {author} {\bibfnamefont {Y.-H.}\ \bibnamefont
  {Zhang}}, \bibinfo {author} {\bibfnamefont {D.}~\bibnamefont {Mao}}, \ and\
  \bibinfo {author} {\bibfnamefont {T.}~\bibnamefont {Senthil}},\ }\bibfield
  {title} {\enquote {\bibinfo {title} {Twisted bilayer graphene aligned with
  hexagonal boron nitride: {A}nomalous {H}all effect and a lattice model},}\
  }\href {\doibase 10.1103/PhysRevResearch.1.033126} {\bibfield  {journal}
  {\bibinfo  {journal} {Phys. Rev. Research}\ }\textbf {\bibinfo {volume}
  {1}},\ \bibinfo {pages} {033126} (\bibinfo {year} {2019})}\BibitemShut
  {NoStop}%
\bibitem [{\citenamefont {Repellin}\ \emph {et~al.}(2020)\citenamefont
  {Repellin}, \citenamefont {Dong}, \citenamefont {Zhang},\ and\ \citenamefont
  {Senthil}}]{Ferromagnetism-in-2020}%
  \BibitemOpen
  \bibfield  {author} {\bibinfo {author} {\bibfnamefont {C.}~\bibnamefont
  {Repellin}}, \bibinfo {author} {\bibfnamefont {Z.}~\bibnamefont {Dong}},
  \bibinfo {author} {\bibfnamefont {Y.-H.}\ \bibnamefont {Zhang}}, \ and\
  \bibinfo {author} {\bibfnamefont {T.}~\bibnamefont {Senthil}},\ }\bibfield
  {title} {\enquote {\bibinfo {title} {{F}erromagnetism in {N}arrow {B}ands of
  {M}oir\'e {S}uperlattices},}\ }\href {\doibase
  10.1103/PhysRevLett.124.187601} {\bibfield  {journal} {\bibinfo  {journal}
  {Phys. Rev. Lett.}\ }\textbf {\bibinfo {volume} {124}},\ \bibinfo {pages}
  {187601} (\bibinfo {year} {2020})}\BibitemShut {NoStop}%
\bibitem [{\citenamefont {Xie}\ and\ \citenamefont
  {MacDonald}(2020)}]{Nature2020xie}%
  \BibitemOpen
  \bibfield  {author} {\bibinfo {author} {\bibfnamefont {M.}~\bibnamefont
  {Xie}}\ and\ \bibinfo {author} {\bibfnamefont {A.~H.}\ \bibnamefont
  {MacDonald}},\ }\bibfield  {title} {\enquote {\bibinfo {title} {Nature of the
  {C}orrelated {I}nsulator {S}tates in {T}wisted {B}ilayer {G}raphene},}\
  }\href {\doibase 10.1103/PhysRevLett.124.097601} {\bibfield  {journal}
  {\bibinfo  {journal} {Phys. Rev. Lett.}\ }\textbf {\bibinfo {volume} {124}},\
  \bibinfo {pages} {097601} (\bibinfo {year} {2020})}\BibitemShut {NoStop}%
\bibitem [{\citenamefont {Bultinck}\ \emph
  {et~al.}(2020{\natexlab{a}})\citenamefont {Bultinck}, \citenamefont {Khalaf},
  \citenamefont {Liu}, \citenamefont {Chatterjee}, \citenamefont {Vishwanath},\
  and\ \citenamefont {Zaletel}}]{Ground-State-2020}%
  \BibitemOpen
  \bibfield  {author} {\bibinfo {author} {\bibfnamefont {N.}~\bibnamefont
  {Bultinck}}, \bibinfo {author} {\bibfnamefont {E.}~\bibnamefont {Khalaf}},
  \bibinfo {author} {\bibfnamefont {S.}~\bibnamefont {Liu}}, \bibinfo {author}
  {\bibfnamefont {S.}~\bibnamefont {Chatterjee}}, \bibinfo {author}
  {\bibfnamefont {A.}~\bibnamefont {Vishwanath}}, \ and\ \bibinfo {author}
  {\bibfnamefont {M.~P.}\ \bibnamefont {Zaletel}},\ }\bibfield  {title}
  {\enquote {\bibinfo {title} {{G}round {S}tate and {H}idden {S}ymmetry of
  {M}agic-{A}ngle {G}raphene at {E}ven {I}nteger {F}illing},}\ }\href {\doibase
  10.1103/PhysRevX.10.031034} {\bibfield  {journal} {\bibinfo  {journal} {Phys.
  Rev. X}\ }\textbf {\bibinfo {volume} {10}},\ \bibinfo {pages} {031034}
  (\bibinfo {year} {2020}{\natexlab{a}})}\BibitemShut {NoStop}%
\bibitem [{\citenamefont {Bultinck}\ \emph
  {et~al.}(2020{\natexlab{b}})\citenamefont {Bultinck}, \citenamefont
  {Chatterjee},\ and\ \citenamefont {Zaletel}}]{Mechanism2020nick}%
  \BibitemOpen
  \bibfield  {author} {\bibinfo {author} {\bibfnamefont {N.}~\bibnamefont
  {Bultinck}}, \bibinfo {author} {\bibfnamefont {S.}~\bibnamefont
  {Chatterjee}}, \ and\ \bibinfo {author} {\bibfnamefont {M.~P.}\ \bibnamefont
  {Zaletel}},\ }\bibfield  {title} {\enquote {\bibinfo {title} {Mechanism for
  {A}nomalous {H}all {F}erromagnetism in {T}wisted {B}ilayer {G}raphene},}\
  }\href {\doibase 10.1103/PhysRevLett.124.166601} {\bibfield  {journal}
  {\bibinfo  {journal} {Phys. Rev. Lett.}\ }\textbf {\bibinfo {volume} {124}},\
  \bibinfo {pages} {166601} (\bibinfo {year} {2020}{\natexlab{b}})}\BibitemShut
  {NoStop}%
\bibitem [{\citenamefont {Zhang}\ \emph {et~al.}(2020)\citenamefont {Zhang},
  \citenamefont {Jiang}, \citenamefont {Wang},\ and\ \citenamefont
  {Zhang}}]{Correlated-insulating-2020}%
  \BibitemOpen
  \bibfield  {author} {\bibinfo {author} {\bibfnamefont {Y.}~\bibnamefont
  {Zhang}}, \bibinfo {author} {\bibfnamefont {K.}~\bibnamefont {Jiang}},
  \bibinfo {author} {\bibfnamefont {Z.}~\bibnamefont {Wang}}, \ and\ \bibinfo
  {author} {\bibfnamefont {F.}~\bibnamefont {Zhang}},\ }\bibfield  {title}
  {\enquote {\bibinfo {title} {{C}orrelated insulating phases of twisted
  bilayer graphene at commensurate filling fractions: {A} {H}artree-{F}ock
  study},}\ }\href {\doibase 10.1103/PhysRevB.102.035136} {\bibfield  {journal}
  {\bibinfo  {journal} {Phys. Rev. B}\ }\textbf {\bibinfo {volume} {102}},\
  \bibinfo {pages} {035136} (\bibinfo {year} {2020})}\BibitemShut {NoStop}%
\bibitem [{\citenamefont {Kang}\ and\ \citenamefont
  {Vafek}(2020)}]{Non-AbelianDirac-2020}%
  \BibitemOpen
  \bibfield  {author} {\bibinfo {author} {\bibfnamefont {J.}~\bibnamefont
  {Kang}}\ and\ \bibinfo {author} {\bibfnamefont {O.}~\bibnamefont {Vafek}},\
  }\bibfield  {title} {\enquote {\bibinfo {title} {{N}on-{A}belian {D}irac node
  braiding and near-degeneracy of correlated phases at odd integer filling in
  magic-angle twisted bilayer graphene},}\ }\href {\doibase
  10.1103/PhysRevB.102.035161} {\bibfield  {journal} {\bibinfo  {journal}
  {Phys. Rev. B}\ }\textbf {\bibinfo {volume} {102}},\ \bibinfo {pages}
  {035161} (\bibinfo {year} {2020})}\BibitemShut {NoStop}%
\bibitem [{\citenamefont {Wu}\ and\ \citenamefont
  {Das~Sarma}(2020)}]{Collective-Excitations-2020}%
  \BibitemOpen
  \bibfield  {author} {\bibinfo {author} {\bibfnamefont {F.}~\bibnamefont
  {Wu}}\ and\ \bibinfo {author} {\bibfnamefont {S.}~\bibnamefont {Das~Sarma}},\
  }\bibfield  {title} {\enquote {\bibinfo {title} {{C}ollective {E}xcitations
  of {Q}uantum {A}nomalous {H}all {F}erromagnets in {T}wisted {B}ilayer
  {G}raphene},}\ }\href {\doibase 10.1103/PhysRevLett.124.046403} {\bibfield
  {journal} {\bibinfo  {journal} {Phys. Rev. Lett.}\ }\textbf {\bibinfo
  {volume} {124}},\ \bibinfo {pages} {046403} (\bibinfo {year}
  {2020})}\BibitemShut {NoStop}%
\bibitem [{\citenamefont {Liu}\ and\ \citenamefont
  {Dai}(2021{\natexlab{b}})}]{Theories-for-2021}%
  \BibitemOpen
  \bibfield  {author} {\bibinfo {author} {\bibfnamefont {J.}~\bibnamefont
  {Liu}}\ and\ \bibinfo {author} {\bibfnamefont {X.}~\bibnamefont {Dai}},\
  }\bibfield  {title} {\enquote {\bibinfo {title} {{T}heories for the
  correlated insulating states and quantum anomalous {H}all effect phenomena in
  twisted bilayer graphene},}\ }\href {\doibase 10.1103/PhysRevB.103.035427}
  {\bibfield  {journal} {\bibinfo  {journal} {Phys. Rev. B}\ }\textbf {\bibinfo
  {volume} {103}},\ \bibinfo {pages} {035427} (\bibinfo {year}
  {2021}{\natexlab{b}})}\BibitemShut {NoStop}%
\bibitem [{\citenamefont {Hejazi}\ \emph {et~al.}(2021)\citenamefont {Hejazi},
  \citenamefont {Chen},\ and\ \citenamefont {Balents}}]{Hybrid-Wannier-2021}%
  \BibitemOpen
  \bibfield  {author} {\bibinfo {author} {\bibfnamefont {K.}~\bibnamefont
  {Hejazi}}, \bibinfo {author} {\bibfnamefont {X.}~\bibnamefont {Chen}}, \ and\
  \bibinfo {author} {\bibfnamefont {L.}~\bibnamefont {Balents}},\ }\bibfield
  {title} {\enquote {\bibinfo {title} {{H}ybrid {W}annier {C}hern bands in
  magic angle twisted bilayer graphene and the quantized anomalous {H}all
  effect},}\ }\href {\doibase 10.1103/PhysRevResearch.3.013242} {\bibfield
  {journal} {\bibinfo  {journal} {Phys. Rev. Research}\ }\textbf {\bibinfo
  {volume} {3}},\ \bibinfo {pages} {013242} (\bibinfo {year}
  {2021})}\BibitemShut {NoStop}%
\bibitem [{\citenamefont {Liu}\ \emph {et~al.}(2021)\citenamefont {Liu},
  \citenamefont {Khalaf}, \citenamefont {Lee},\ and\ \citenamefont
  {Vishwanath}}]{Nematic-topological-2021}%
  \BibitemOpen
  \bibfield  {author} {\bibinfo {author} {\bibfnamefont {S.}~\bibnamefont
  {Liu}}, \bibinfo {author} {\bibfnamefont {E.}~\bibnamefont {Khalaf}},
  \bibinfo {author} {\bibfnamefont {J.~Y.}\ \bibnamefont {Lee}}, \ and\
  \bibinfo {author} {\bibfnamefont {A.}~\bibnamefont {Vishwanath}},\ }\bibfield
   {title} {\enquote {\bibinfo {title} {{N}ematic topological semimetal and
  insulator in magic-angle bilayer graphene at charge neutrality},}\ }\href
  {\doibase 10.1103/PhysRevResearch.3.013033} {\bibfield  {journal} {\bibinfo
  {journal} {Phys. Rev. Research}\ }\textbf {\bibinfo {volume} {3}},\ \bibinfo
  {pages} {013033} (\bibinfo {year} {2021})}\BibitemShut {NoStop}%
\bibitem [{\citenamefont {Da~Liao}\ \emph {et~al.}(2021)\citenamefont
  {Da~Liao}, \citenamefont {Kang}, \citenamefont {Brei\o{}}, \citenamefont
  {Xu}, \citenamefont {Wu}, \citenamefont {Andersen}, \citenamefont
  {Fernandes},\ and\ \citenamefont
  {Meng}}]{Correlation-InducedInsulating-2021}%
  \BibitemOpen
  \bibfield  {author} {\bibinfo {author} {\bibfnamefont {Y.}~\bibnamefont
  {Da~Liao}}, \bibinfo {author} {\bibfnamefont {J.}~\bibnamefont {Kang}},
  \bibinfo {author} {\bibfnamefont {C.~N.}\ \bibnamefont {Brei\o{}}}, \bibinfo
  {author} {\bibfnamefont {X.~Y.}\ \bibnamefont {Xu}}, \bibinfo {author}
  {\bibfnamefont {H.-Q.}\ \bibnamefont {Wu}}, \bibinfo {author} {\bibfnamefont
  {B.~M.}\ \bibnamefont {Andersen}}, \bibinfo {author} {\bibfnamefont {R.~M.}\
  \bibnamefont {Fernandes}}, \ and\ \bibinfo {author} {\bibfnamefont {Z.~Y.}\
  \bibnamefont {Meng}},\ }\bibfield  {title} {\enquote {\bibinfo {title}
  {{C}orrelation-{I}nduced {I}nsulating {T}opological {P}hases at {C}harge
  {N}eutrality in {T}wisted {B}ilayer {G}raphene},}\ }\href {\doibase
  10.1103/PhysRevX.11.011014} {\bibfield  {journal} {\bibinfo  {journal} {Phys.
  Rev. X}\ }\textbf {\bibinfo {volume} {11}},\ \bibinfo {pages} {011014}
  (\bibinfo {year} {2021})}\BibitemShut {NoStop}%
\bibitem [{\citenamefont {Lin}\ and\ \citenamefont
  {Ni}(2020)}]{Symmetry-breaking-2020lin}%
  \BibitemOpen
  \bibfield  {author} {\bibinfo {author} {\bibfnamefont {X.}~\bibnamefont
  {Lin}}\ and\ \bibinfo {author} {\bibfnamefont {J.}~\bibnamefont {Ni}},\
  }\bibfield  {title} {\enquote {\bibinfo {title} {Symmetry breaking in the
  double moir\'e superlattices of relaxed twisted bilayer graphene on hexagonal
  boron nitride},}\ }\href {\doibase 10.1103/PhysRevB.102.035441} {\bibfield
  {journal} {\bibinfo  {journal} {Phys. Rev. B}\ }\textbf {\bibinfo {volume}
  {102}},\ \bibinfo {pages} {035441} (\bibinfo {year} {2020})}\BibitemShut
  {NoStop}%
\bibitem [{\citenamefont {Bernevig}\ \emph
  {et~al.}(2021{\natexlab{a}})\citenamefont {Bernevig}, \citenamefont {Song},
  \citenamefont {Regnault},\ and\ \citenamefont
  {Lian}}]{Twisted-bilayeriii-2021}%
  \BibitemOpen
  \bibfield  {author} {\bibinfo {author} {\bibfnamefont {B.~A.}\ \bibnamefont
  {Bernevig}}, \bibinfo {author} {\bibfnamefont {Z.-D.}\ \bibnamefont {Song}},
  \bibinfo {author} {\bibfnamefont {N.}~\bibnamefont {Regnault}}, \ and\
  \bibinfo {author} {\bibfnamefont {B.}~\bibnamefont {Lian}},\ }\bibfield
  {title} {\enquote {\bibinfo {title} {{T}wisted bilayer graphene. {I}{I}{I}.
  {I}nteracting {H}amiltonian and exact symmetries},}\ }\href {\doibase
  10.1103/PhysRevB.103.205413} {\bibfield  {journal} {\bibinfo  {journal}
  {Phys. Rev. B}\ }\textbf {\bibinfo {volume} {103}},\ \bibinfo {pages}
  {205413} (\bibinfo {year} {2021}{\natexlab{a}})}\BibitemShut {NoStop}%
\bibitem [{\citenamefont {Lian}\ \emph {et~al.}(2021)\citenamefont {Lian},
  \citenamefont {Song}, \citenamefont {Regnault}, \citenamefont {Efetov},
  \citenamefont {Yazdani},\ and\ \citenamefont
  {Bernevig}}]{Twisted-bilayeriv-2021}%
  \BibitemOpen
  \bibfield  {author} {\bibinfo {author} {\bibfnamefont {B.}~\bibnamefont
  {Lian}}, \bibinfo {author} {\bibfnamefont {Z.-D.}\ \bibnamefont {Song}},
  \bibinfo {author} {\bibfnamefont {N.}~\bibnamefont {Regnault}}, \bibinfo
  {author} {\bibfnamefont {D.~K.}\ \bibnamefont {Efetov}}, \bibinfo {author}
  {\bibfnamefont {A.}~\bibnamefont {Yazdani}}, \ and\ \bibinfo {author}
  {\bibfnamefont {B.~A.}\ \bibnamefont {Bernevig}},\ }\bibfield  {title}
  {\enquote {\bibinfo {title} {{T}wisted bilayer graphene. {I}{V}. {E}xact
  insulator ground states and phase diagram},}\ }\href {\doibase
  10.1103/PhysRevB.103.205414} {\bibfield  {journal} {\bibinfo  {journal}
  {Phys. Rev. B}\ }\textbf {\bibinfo {volume} {103}},\ \bibinfo {pages}
  {205414} (\bibinfo {year} {2021})}\BibitemShut {NoStop}%
\bibitem [{\citenamefont {Bernevig}\ \emph
  {et~al.}(2021{\natexlab{b}})\citenamefont {Bernevig}, \citenamefont {Lian},
  \citenamefont {Cowsik}, \citenamefont {Xie}, \citenamefont {Regnault},\ and\
  \citenamefont {Song}}]{Twisted-bilayerv-2021}%
  \BibitemOpen
  \bibfield  {author} {\bibinfo {author} {\bibfnamefont {B.~A.}\ \bibnamefont
  {Bernevig}}, \bibinfo {author} {\bibfnamefont {B.}~\bibnamefont {Lian}},
  \bibinfo {author} {\bibfnamefont {A.}~\bibnamefont {Cowsik}}, \bibinfo
  {author} {\bibfnamefont {F.}~\bibnamefont {Xie}}, \bibinfo {author}
  {\bibfnamefont {N.}~\bibnamefont {Regnault}}, \ and\ \bibinfo {author}
  {\bibfnamefont {Z.-D.}\ \bibnamefont {Song}},\ }\bibfield  {title} {\enquote
  {\bibinfo {title} {{T}wisted bilayer graphene. {V}. {E}xact analytic
  many-body excitations in {C}oulomb {H}amiltonians: {C}harge gap, {G}oldstone
  modes, and absence of {C}ooper pairing},}\ }\href {\doibase
  10.1103/PhysRevB.103.205415} {\bibfield  {journal} {\bibinfo  {journal}
  {Phys. Rev. B}\ }\textbf {\bibinfo {volume} {103}},\ \bibinfo {pages}
  {205415} (\bibinfo {year} {2021}{\natexlab{b}})}\BibitemShut {NoStop}%
\bibitem [{\citenamefont {Kwan}\ \emph {et~al.}(2021)\citenamefont {Kwan},
  \citenamefont {Wagner}, \citenamefont {Soejima}, \citenamefont {Zaletel},
  \citenamefont {Simon}, \citenamefont {Parameswaran},\ and\ \citenamefont
  {Bultinck}}]{Kekul-Spiral-2021}%
  \BibitemOpen
  \bibfield  {author} {\bibinfo {author} {\bibfnamefont {Y.~H.}\ \bibnamefont
  {Kwan}}, \bibinfo {author} {\bibfnamefont {G.}~\bibnamefont {Wagner}},
  \bibinfo {author} {\bibfnamefont {T.}~\bibnamefont {Soejima}}, \bibinfo
  {author} {\bibfnamefont {M.~P.}\ \bibnamefont {Zaletel}}, \bibinfo {author}
  {\bibfnamefont {S.~H.}\ \bibnamefont {Simon}}, \bibinfo {author}
  {\bibfnamefont {S.~A.}\ \bibnamefont {Parameswaran}}, \ and\ \bibinfo
  {author} {\bibfnamefont {N.}~\bibnamefont {Bultinck}},\ }\bibfield  {title}
  {\enquote {\bibinfo {title} {{K}ekul\'e {S}piral {O}rder at {A}ll {N}onzero
  {I}nteger {F}illings in {T}wisted {B}ilayer {G}raphene},}\ }\href {\doibase
  10.1103/PhysRevX.11.041063} {\bibfield  {journal} {\bibinfo  {journal} {Phys.
  Rev. X}\ }\textbf {\bibinfo {volume} {11}},\ \bibinfo {pages} {041063}
  (\bibinfo {year} {2021})}\BibitemShut {NoStop}%
\bibitem [{\citenamefont {Cea}\ \emph {et~al.}(2020)\citenamefont {Cea},
  \citenamefont {Pantale\'on},\ and\ \citenamefont
  {Guinea}}]{Band-structure-2020}%
  \BibitemOpen
  \bibfield  {author} {\bibinfo {author} {\bibfnamefont {T.}~\bibnamefont
  {Cea}}, \bibinfo {author} {\bibfnamefont {P.~A.}\ \bibnamefont
  {Pantale\'on}}, \ and\ \bibinfo {author} {\bibfnamefont {F.}~\bibnamefont
  {Guinea}},\ }\bibfield  {title} {\enquote {\bibinfo {title} {Band structure
  of twisted bilayer graphene on hexagonal boron nitride},}\ }\href {\doibase
  10.1103/PhysRevB.102.155136} {\bibfield  {journal} {\bibinfo  {journal}
  {Phys. Rev. B}\ }\textbf {\bibinfo {volume} {102}},\ \bibinfo {pages}
  {155136} (\bibinfo {year} {2020})}\BibitemShut {NoStop}%
\bibitem [{\citenamefont {Lin}\ \emph {et~al.}(2021)\citenamefont {Lin},
  \citenamefont {Su},\ and\ \citenamefont {Ni}}]{Misalignment2021Lin}%
  \BibitemOpen
  \bibfield  {author} {\bibinfo {author} {\bibfnamefont {X.}~\bibnamefont
  {Lin}}, \bibinfo {author} {\bibfnamefont {K.}~\bibnamefont {Su}}, \ and\
  \bibinfo {author} {\bibfnamefont {J.}~\bibnamefont {Ni}},\ }\bibfield
  {title} {\enquote {\bibinfo {title} {Misalignment instability in magic-angle
  twisted bilayer graphene on hexagonal boron nitride},}\ }\href {\doibase
  10.1088/2053-1583/abddcb} {\bibfield  {journal} {\bibinfo  {journal} {2D
  Mater.}\ }\textbf {\bibinfo {volume} {8}},\ \bibinfo {pages} {025025}
  (\bibinfo {year} {2021})}\BibitemShut {NoStop}%
\bibitem [{\citenamefont {Shi}\ \emph {et~al.}(2021)\citenamefont {Shi},
  \citenamefont {Zhu},\ and\ \citenamefont {MacDonald}}]{Moire2021Shi}%
  \BibitemOpen
  \bibfield  {author} {\bibinfo {author} {\bibfnamefont {J.}~\bibnamefont
  {Shi}}, \bibinfo {author} {\bibfnamefont {J.}~\bibnamefont {Zhu}}, \ and\
  \bibinfo {author} {\bibfnamefont {A.~H.}\ \bibnamefont {MacDonald}},\
  }\bibfield  {title} {\enquote {\bibinfo {title} {Moir\'e commensurability and
  the quantum anomalous {H}all effect in twisted bilayer graphene on hexagonal
  boron nitride},}\ }\href {\doibase 10.1103/PhysRevB.103.075122} {\bibfield
  {journal} {\bibinfo  {journal} {Phys. Rev. B}\ }\textbf {\bibinfo {volume}
  {103}},\ \bibinfo {pages} {075122} (\bibinfo {year} {2021})}\BibitemShut
  {NoStop}%
\bibitem [{\citenamefont {Mao}\ and\ \citenamefont
  {Senthil}(2021)}]{Quasiperiodicity2021Mao}%
  \BibitemOpen
  \bibfield  {author} {\bibinfo {author} {\bibfnamefont {D.}~\bibnamefont
  {Mao}}\ and\ \bibinfo {author} {\bibfnamefont {T.}~\bibnamefont {Senthil}},\
  }\bibfield  {title} {\enquote {\bibinfo {title} {Quasiperiodicity, band
  topology, and moir\'e graphene},}\ }\href {\doibase
  10.1103/PhysRevB.103.115110} {\bibfield  {journal} {\bibinfo  {journal}
  {Phys. Rev. B}\ }\textbf {\bibinfo {volume} {103}},\ \bibinfo {pages}
  {115110} (\bibinfo {year} {2021})}\BibitemShut {NoStop}%
\bibitem [{\citenamefont {Shin}\ \emph {et~al.}(2021)\citenamefont {Shin},
  \citenamefont {Park}, \citenamefont {Chittari}, \citenamefont {Sun},\ and\
  \citenamefont {Jung}}]{Electron-hole-asymmetry-2021}%
  \BibitemOpen
  \bibfield  {author} {\bibinfo {author} {\bibfnamefont {J.}~\bibnamefont
  {Shin}}, \bibinfo {author} {\bibfnamefont {Y.}~\bibnamefont {Park}}, \bibinfo
  {author} {\bibfnamefont {B.~L.}\ \bibnamefont {Chittari}}, \bibinfo {author}
  {\bibfnamefont {J.-H.}\ \bibnamefont {Sun}}, \ and\ \bibinfo {author}
  {\bibfnamefont {J.}~\bibnamefont {Jung}},\ }\bibfield  {title} {\enquote
  {\bibinfo {title} {{E}lectron-hole asymmetry and band gaps of commensurate
  double moire patterns in twisted bilayer graphene on hexagonal boron
  nitride},}\ }\href {\doibase 10.1103/PhysRevB.103.075423} {\bibfield
  {journal} {\bibinfo  {journal} {Phys. Rev. B}\ }\textbf {\bibinfo {volume}
  {103}},\ \bibinfo {pages} {075423} (\bibinfo {year} {2021})}\BibitemShut
  {NoStop}%
\bibitem [{\citenamefont {Long}\ \emph {et~al.}(2022)\citenamefont {Long},
  \citenamefont {Pantale\'on}, \citenamefont {Zhan}, \citenamefont {Guinea},
  \citenamefont {Silva-Guill\'en},\ and\ \citenamefont
  {Yuan}}]{An-atomistic-2022}%
  \BibitemOpen
  \bibfield  {author} {\bibinfo {author} {\bibfnamefont {M.}~\bibnamefont
  {Long}}, \bibinfo {author} {\bibfnamefont {P.~A.}\ \bibnamefont
  {Pantale\'on}}, \bibinfo {author} {\bibfnamefont {Z.}~\bibnamefont {Zhan}},
  \bibinfo {author} {\bibfnamefont {F.}~\bibnamefont {Guinea}}, \bibinfo
  {author} {\bibfnamefont {J.~Ángel}\ \bibnamefont {Silva-Guill\'en}}, \ and\
  \bibinfo {author} {\bibfnamefont {S.}~\bibnamefont {Yuan}},\ }\bibfield
  {title} {\enquote {\bibinfo {title} {{A}n atomistic approach for the
  structural and electronic properties of twisted bilayer graphene-boron
  nitride heterostructures},}\ }\href {\doibase 10.1038/s41524-022-00763-1}
  {\bibfield  {journal} {\bibinfo  {journal} {npj Computational Materials}\
  }\textbf {\bibinfo {volume} {8}},\ \bibinfo {pages} {73} (\bibinfo {year}
  {2022})}\BibitemShut {NoStop}%
\bibitem [{\citenamefont {Serlin}\ \emph {et~al.}(2020)\citenamefont {Serlin},
  \citenamefont {Tschirhart}, \citenamefont {Polshyn}, \citenamefont {Zhang},
  \citenamefont {Zhu}, \citenamefont {Watanabe}, \citenamefont {Taniguchi},
  \citenamefont {Balents},\ and\ \citenamefont {Young}}]{Intrinsic2020Serlin}%
  \BibitemOpen
  \bibfield  {author} {\bibinfo {author} {\bibfnamefont {M.}~\bibnamefont
  {Serlin}}, \bibinfo {author} {\bibfnamefont {C.~L.}\ \bibnamefont
  {Tschirhart}}, \bibinfo {author} {\bibfnamefont {H.}~\bibnamefont {Polshyn}},
  \bibinfo {author} {\bibfnamefont {Y.}~\bibnamefont {Zhang}}, \bibinfo
  {author} {\bibfnamefont {J.}~\bibnamefont {Zhu}}, \bibinfo {author}
  {\bibfnamefont {K.}~\bibnamefont {Watanabe}}, \bibinfo {author}
  {\bibfnamefont {T.}~\bibnamefont {Taniguchi}}, \bibinfo {author}
  {\bibfnamefont {L.}~\bibnamefont {Balents}}, \ and\ \bibinfo {author}
  {\bibfnamefont {A.~F.}\ \bibnamefont {Young}},\ }\bibfield  {title} {\enquote
  {\bibinfo {title} {Intrinsic quantized anomalous {H}all effect in a moir{\'e}
  heterostructure},}\ }\href {\doibase 10.1126/science.aay5533} {\bibfield
  {journal} {\bibinfo  {journal} {Science}\ }\textbf {\bibinfo {volume}
  {367}},\ \bibinfo {pages} {900} (\bibinfo {year} {2020})}\BibitemShut
  {NoStop}%
\bibitem [{\citenamefont {Tschirhart}\ \emph {et~al.}(2021)\citenamefont
  {Tschirhart}, \citenamefont {Serlin}, \citenamefont {Polshyn}, \citenamefont
  {Shragai}, \citenamefont {Xia}, \citenamefont {Zhu}, \citenamefont {Zhang},
  \citenamefont {Watanabe}, \citenamefont {Taniguchi}, \citenamefont {Huber},\
  and\ \citenamefont {Young}}]{Imaging-orbital-2021}%
  \BibitemOpen
  \bibfield  {author} {\bibinfo {author} {\bibfnamefont {C.~L.}\ \bibnamefont
  {Tschirhart}}, \bibinfo {author} {\bibfnamefont {M.}~\bibnamefont {Serlin}},
  \bibinfo {author} {\bibfnamefont {H.}~\bibnamefont {Polshyn}}, \bibinfo
  {author} {\bibfnamefont {A.}~\bibnamefont {Shragai}}, \bibinfo {author}
  {\bibfnamefont {Z.}~\bibnamefont {Xia}}, \bibinfo {author} {\bibfnamefont
  {J.}~\bibnamefont {Zhu}}, \bibinfo {author} {\bibfnamefont {Y.}~\bibnamefont
  {Zhang}}, \bibinfo {author} {\bibfnamefont {K.}~\bibnamefont {Watanabe}},
  \bibinfo {author} {\bibfnamefont {T.}~\bibnamefont {Taniguchi}}, \bibinfo
  {author} {\bibfnamefont {M.~E.}\ \bibnamefont {Huber}}, \ and\ \bibinfo
  {author} {\bibfnamefont {A.~F.}\ \bibnamefont {Young}},\ }\bibfield  {title}
  {\enquote {\bibinfo {title} {{I}maging orbital ferromagnetism in a moir\'e
  {C}hern insulator},}\ }\href {\doibase 10.1126/science.abd3190} {\bibfield
  {journal} {\bibinfo  {journal} {Science}\ }\textbf {\bibinfo {volume}
  {372}},\ \bibinfo {pages} {1323} (\bibinfo {year} {2021})}\BibitemShut
  {NoStop}%
\bibitem [{\citenamefont {Kumar}\ \emph {et~al.}(2021)\citenamefont {Kumar},
  \citenamefont {Xie},\ and\ \citenamefont
  {MacDonald}}]{Lattice-collective-2021}%
  \BibitemOpen
  \bibfield  {author} {\bibinfo {author} {\bibfnamefont {A.}~\bibnamefont
  {Kumar}}, \bibinfo {author} {\bibfnamefont {M.}~\bibnamefont {Xie}}, \ and\
  \bibinfo {author} {\bibfnamefont {A.~H.}\ \bibnamefont {MacDonald}},\
  }\bibfield  {title} {\enquote {\bibinfo {title} {{L}attice collective modes
  from a continuum model of magic-angle twisted bilayer graphene},}\ }\href
  {\doibase 10.1103/PhysRevB.104.035119} {\bibfield  {journal} {\bibinfo
  {journal} {Phys. Rev. B}\ }\textbf {\bibinfo {volume} {104}},\ \bibinfo
  {pages} {035119} (\bibinfo {year} {2021})}\BibitemShut {NoStop}%
\end{thebibliography}

%

\end{document}